\documentclass[10pt,twocolumn]{IEEEtran}

\IEEEoverridecommandlockouts                              % This command is only
\usepackage{amsmath,graphicx,amsfonts,amssymb,amsthm,epsfig,mathrsfs,balance,epsfig}
\usepackage{subcaption}
\usepackage{tikz}
\usetikzlibrary{shapes}
\usepackage{color,multirow,rotating}
\usepackage{stmaryrd}
\usepackage{algorithm,algpseudocode,algorithmicx}
\usepackage{cite,url,framed,bm,soul}

\newtheorem{theorem}{Theorem}
\newtheorem{corollary}[theorem]{Corollary}

\newtheorem{proposition}{Proposition}
\newtheorem{remark}{Remark}

\renewcommand{\det}{{\mathrm{det}}}
\newcommand{\tr}{{\mathrm{trace}}}

\newcommand{\vol}{{\mathrm{vol}}}
\newcommand{\blkdiag}{{\mathrm{blkdiag}}}
\newcommand{\dist}{{\mathrm{dist}}}

\newcommand{\differential}{{\rm{d}}}

\renewcommand{\qed}{\hfill\ensuremath{\blacksquare}}

\allowdisplaybreaks

\title{\LARGE \bf
The Curious Case of Integrator Reach Sets, Part I: Basic Theory
}

\author{Shadi Haddad, Abhishek Halder, \emph{Senior Member, IEEE}% <-this % stops a space
\thanks{Shadi Haddad and Abhishek Halder are with the Department of Applied Mathematics, University of California, Santa Cruz, CA 95064, USA,
        {\tt\small{\{shhaddad,ahalder\}@ucsc.edu}}%
}}

\begin{document}

\maketitle
\thispagestyle{empty}
\pagestyle{empty}

%%%%%%%%%%%%%%%%%%%%%%%%%%%%%%%%%%%%%%%%%%%%%%%%%%%%%%%%%%%%%%%%%%%%%%%%%%%%%%%%
\begin{abstract}
This is the first of a two part paper investigating the geometry of the integrator reach sets, and the applications thereof. In this Part I, assuming box-valued input uncertainties, we establish that this compact convex reach set is semialgebraic, translated zonoid, and not a spectrahedron. We derive the parametric as well as the implicit representation of the boundary of this reach set. We also deduce the closed form formula for the volume and diameter of this set, and discuss their scaling with state dimension and time. We point out that these results may be utilized in benchmarking the performance of the reach set over-approximation algorithms.
\end{abstract}

%%%%%%%%%%%%%%%%%%%%%%%%%%%%%%%%%%%%%%%%%%%%%%%%%%%%%%%%%%%%%%%%%%%%%%%%%%%%%%%%


%%%%%%%%%%%%%%%%%%%%%%%%%%%%%%%%%%%%%%%%%%%%%%%%%%%%%%%%%%%%%%%%%%%%%%%%%%%%%%%%%%%%%%%%%%%%%%%%%%%%%%%%%%%%%%%%%%%%%%%%%%%%%%%%%%%%%%%%%%%%%%%%%%%%%%%%%%%%%%%

\vspace*{-0.1in}

\section{Introduction}\label{sec:Intro}
Integrators with bounded controls are ubiquitous in systems-control. They appear as Brunovsky normal forms for the feedback linearizable nonlinear systems. They also appear frequently as benchmark problems to demonstrate the performance of the reach set computation algorithms. Despite their prominence, specific results on the geometry of the integrator reach sets are not available in the systems-control literature. Broadly speaking, the existing results come in two flavors. On one hand, very generic statements are known, e.g., these reach sets are compact convex sets whenever the set of initial conditions is compact convex, and the controls take values from a compact (not necessarily convex) set \cite{varaiya2000reach}. On the other hand, several numerical toolboxes \cite{althoff2015introduction,kurzhanskiy2006ellipsoidal} are available for tight outer approximation of the reach sets over computationally benign geometric families such as ellipsoids and zonotopes. The lack of concrete geometric results imply the absence of ground truth when comparing the efficacy of different algorithms, and one has to content with graphical or statistical (e.g., Monte Carlo) comparisons.

Building on the preliminary results in \cite{haddad2020convex}, this paper undertakes a systematic study of the integrator reach sets. In particular, we answer the following basic questions:

\begin{itemize}

\item[\textbf{Q1.}] what kind of compact convex sets are these (Section \ref{sec:Taxonomy})?

\item[\textbf{Q2.}] how big are these sets (Section \ref{sec:Size})?

\item[\textbf{Q3.}] how these results on the geometry of integrator reach sets can be applied in practice (Section \ref{sec:Applications})?
	
\end{itemize}
%In resolving these questions, we uncover a rich interplay between algebra and convex geometry.  

We consider the integrator dynamics having $d$ states and $m$ inputs with relative degree vector $\bm{r} = \left(r_1,r_2,\hdots,r_m \right)^{\top} \in \mathbb{Z}_{+}^{m}$ (vector of positive integers). In other words, we consider a Brunovsky normal form with $m$ integrators where the $j$th integrator has degree $r_{j}$ for $j\in[m]:=\{1,\hdots,m\}$. The dynamics is %given by
\begin{align}
\dot{\bm{x}} = \bm{A}\bm{x} + \bm{B}\bm{u}, \quad\bm{x}\in\mathbb{R}^{d}, \quad \bm{u}(\cdot)\in\mathcal{U}\subset\mathbb{R}^{m},
\label{IntegratorDyn}
\end{align}
where $r_1 + r_2 + ... + r_m = d$, the set $\mathcal{U}$ is compact, and
\begin{subequations}
\begin{align}
&\!\!\bm{A}\! :=\!\blkdiag\!\left(\bm{A}_{1}, ..., \bm{A}_{m}\right),\quad \!\!\bm{B}\! :=\!\blkdiag\!\left(\bm{b}_{1} , ...,  \bm{b}_{m}\right),\label{blkdiagAB}\\
&\!\!\bm{A}_{j} := \left(\bm{0}_{r_{j}\times 1}~\bm{|}~\bm{e}_{1}^{r_{j}}\bm{|}~\bm{e}_2^{r_{j}}\bm{|} ... \bm{|}~\boldsymbol{e}_{r_{j} - 1}^{r_{j}}\right), \qquad\!\!\!\bm{b}_{j} :=  \bm{e}_{r_j}^{r_j}.\label{defBlocks}
\end{align}
\label{defAB}
\end{subequations}
In (\ref{blkdiagAB}), the symbol $\blkdiag\left(\cdot\right)$ denotes a block diagonal matrix whose arguments constitute its diagonal blocks. In (\ref{defBlocks}), the notation $\bm{0}_{r_{j}\times 1} $ stands for the ${r_{j}\times 1} $ column vector of zeros, and $\bm{e}_{k}^{\ell}$ denotes the $k$th basis (column) vector in $\mathbb{R}^{\ell}$ for $k\leq \ell$. The symbol $(...| ... | ...)$ denotes horizontal concatenation. 

Notice that an integrator that replaces the ones appearing in the system matrices with arbitrary nonzero reals, is always reducible to the normal form \eqref{IntegratorDyn}-\eqref{defAB} by renaming the variables. For instance, for any $a,b\in\mathbb{R}\setminus\{0\}$, the system $\dot{x}_{1} = a x_{2},  \dot{x}_{2} = b u_{1},$
is equivalent to $\dot{x}_{1} = \tilde{x}_{2}, \dot{\tilde{x}}_{2} = \tilde{u}_{1},$ where $\tilde{x}_{2} := a x_{2}$, $\tilde{u}_{1} := ab u_{1}$.

Let $\mathcal{R}\left(\mathcal{X}_{0},t\right)$ denote the \emph{forward reach set} of (\ref{IntegratorDyn}) at time $t>0$, starting from a given compact convex set of initial conditions $\mathcal{X}_{0}\subset\mathbb{R}^{d}$, i.e.,
\begin{align}
\mathcal{R}&\left(\mathcal{X}_{0},t\right) := \!\!\!\bigcup_{\text{measurable}\:\bm{u}(\cdot)\in\mathcal{U}\subset\mathbb{R}^{m}}\!\!\!\big\{\bm{x}(t)\in\mathbb{R}^{d} \mid \text{\eqref{IntegratorDyn} and \eqref{defAB} hold},\nonumber\\
&\qquad\bm{x}(t=0) \in\mathcal{X}_{0}\;\text{compact convex}, \quad \mathcal{U}\;\text{compact}\big\}.
\label{DefReachSet}	
\end{align}
In words, $\mathcal{R}\left(\mathcal{X}_{0},t\right)$ is the set of all states that the controlled dynamics \eqref{IntegratorDyn}-\eqref{defAB} can reach at time $t>0$, starting from the set $\mathcal{X}_{0}$ at $t=0$, with measurable control $\bm{u}(\cdot)\in\mathcal{U}$ compact. Formally,
\begin{align}
\!\!\mathcal{R}\left(\mathcal{X}_{0},t\right)\! &= \exp(t\bm{A})\mathcal{X}_{0} \dotplus \!\!\int_{0}^{t}\!\!\!\exp\left((t-\tau)\bm{A}\right)\bm{B}\mathcal{U}\:\differential\tau\nonumber\\
&= \exp(t\bm{A})\mathcal{X}_{0} \dotplus \!\!\int_{0}^{t}\!\!\!\exp\left(s\bm{A}\right)\bm{B}\mathcal{U}\:\differential s,
\label{SetValuedIntegral}	
\end{align}
where $\dotplus$ denotes the Minkowski sum. The set-valued integral \cite{aumann1965integrals} in (\ref{SetValuedIntegral}) is defined for any point-to-set function $F(\cdot)$, as 
\begin{align}
\int_{0}^{t}F(s)\differential s := \lim_{\Delta\downarrow 0}\:\sum_{i=0}^{\lfloor t/\Delta\rfloor}\Delta F(i\Delta),
\label{DefIntegralOfSetValuedFn}
\end{align}
where the summation symbol $\Sigma$ denotes the Minkowski sum, and $\lfloor\cdot\rfloor$ is the floor operator; see e.g., \cite{varaiya2000reach}. %We will often consider the special case of singleton initial set $\mathcal{X}_{0}\equiv\{\bm{x}_{0}\}$.
Our objective is to study the geometry of (\ref{SetValuedIntegral}) in detail.

This paper significantly expands our preliminary works \cite{haddad2020convex,haddad2022boundary}: here we consider multi-input integrators as opposed to the single input case considered in \cite{haddad2020convex}. Even for the single input case, while \cite[Thm. 1]{haddad2020convex} derived an exact formula for the volume of the reach set, that formula involved limit and nested sums, and in that sense, was not really a closed-form formula \cite{borwein2013closed} -- certainly not amenable for numerical computation. In this paper, we derive closed-form formula for the general multi-input case when the input uncertainty is box-valued, i.e., $\mathcal{U}$ is hyperrectangle. In the same setting, the present paper addresses previously unexplored directions: the scaling laws for the volume and diameter of integrator reach sets, exact parametric and implicit equations for the boundary, and the classification of these sets. 

The paper is structured as follows. After reviewing some preliminary concepts in Sec. \ref{subsec:prelim}, we consider the integrator reach set resulting from box-valued input set uncertainty in Sec. \ref{sec:BoxValued}. The results on taxonomy and the boundary of the corresponding reach set are provided in Sec. \ref{sec:Taxonomy}. The results on the size of this set are collected in Sec. \ref{sec:Size}. The application of these results for benchmarking the reach set over-approximation algorithms are discussed in Sec. \ref{sec:Applications}. {{All proofs are deferred to the Appendix.}}  
 Sec. \ref{sec:Epilogue} summarizes the paper, and outlines the directions pursued in its sequel Part II.

\section{Preliminaries}\label{subsec:prelim}
In the following, we summarize some preliminaries which will be useful in the main body and in the Appendix.
\subsubsection{State transition matrix}\label{subsubsec:STM}
For $0\leq s < t$, the state transition matrix $\bm{\Phi}(t,s)$ associated with (\ref{defAB}) is
\begin{align*}
\bm{\Phi}(t,s)&\equiv\exp(\bm{A}(t-s)) \\
&= \blkdiag\left(\exp(\bm{A}_{1}(t-s)),\hdots,\exp(\bm{A}_{m}(t-s))\right),
\end{align*}
with each diagonal block {{is}} upper triangular. {{Specifically, the $j$th diagonal block of size $r_{j}\times r_{j}$ is written element-wise as}}
\begin{align}
\exp(\bm{A}_{j}(t-s)) := \begin{cases}
 \dfrac{(t-s)^{\ell-k}}{(\ell-k)!} & \text{for}\; k\leq\ell,\\
 0 & \text{otherwise},	
 \end{cases}
\label{diagBlocksOfSTM}	
\end{align}
{{where $k$ is the row index, $\ell$ is the column index, and $k,\ell\in[r_{j}]$ for each $j\in[m]$.}} The diagonal entries in (\ref{diagBlocksOfSTM}) are unity.

\subsubsection{Support function}\label{subsubsec:sptfn}
 The support function $h_{\mathcal{K}}(\cdot)$ of a compact convex set $\mathcal{K} \subset \mathbb{R}^{d}$, is given by
\begin{align}
h_{\mathcal{K}}(\bm{y}) := \underset{\bm{x}\in\mathcal{K}}{\sup}~\langle\bm{y},\bm{x}\rangle, \quad \bm{y}\in\mathbb{R}^{d},
\label{DefSptFn}	
\end{align}
where $\langle\cdot,\cdot\rangle$ denotes the standard Euclidean inner product. Geometrically, $h_{\mathcal{K}}(\bm{y})$ gives the signed distance of the supporting hyperplane of $\mathcal{K}$ with outer normal vector $\bm{y}$, measured from the origin. Furthermore, the supporting hyperplane at $\bm{x}^{\text{bdy}}\in\partial\mathcal{K}$ is $\langle\bm{y},\bm{x}^{\text{bdy}}\rangle = h_{\mathcal{K}}(\bm{y})$, and we can write
\[\mathcal{K} = \big\{\bm{x}\in\mathbb{R}^{d} \mid \langle\bm{y},\bm{x}\rangle \leq h_{\mathcal{K}}(\bm{y})\;\text{for all}\;\bm{y}\in\mathbb{R}^{d}\big\}.\]
For compact $\mathcal{K}_1,\mathcal{K}_2\subset\mathbb{R}^{d}$, 
\begin{align}
\mathcal{K}_1\subseteq\mathcal{K}_2 \quad\text{if and only if}\quad h_{\mathcal{K}_1}(\cdot)\leq h_{\mathcal{K}_2}(\cdot).
\label{SptFnInequality}	
\end{align}
The support function $h_{\mathcal{K}}\left(\bm{y}\right)$ is convex in $\bm{y}$. For more details on the support function, we refer the readers to \cite[Ch. V]{hiriart2013convex}.

The support function $h_{\mathcal{K}}(\bm{y})$ uniquely determines the set $\mathcal{K}$. Given matrix-vector pair $(\bm{\Gamma},\bm{\gamma})\in\mathbb{R}^{d\times d}\times\mathbb{R}^{d}$, the support function of the affine transform $\bm{\Gamma}\mathcal{K}+\bm{\gamma}$ is 
\begin{align}
h_{\bm{\Gamma}\mathcal{K}+\bm{\gamma}}(\bm{y}) = \langle\bm{y},\bm{\gamma}\rangle + h_{\mathcal{K}}(\bm{\Gamma}^{\top}\bm{y}).
\label{SptFnAffineTransform}	
\end{align}

Given a function $f:\mathbb{R}^{d}\mapsto\mathbb{R}\cup\{+\infty\}$, its Legendre-Fenchel conjugate is 
\begin{align}
f^{*}(\bm{y}) := \underset{\bm{x}\in\operatorname{domain}(f)}{\sup}\:\{\langle\bm{y},\bm{x}\rangle - f(\bm{x})\}, \quad \bm{y}\in\mathbb{R}^{d}.
\label{LegFenchelConjugate}	
\end{align}
From (\ref{DefSptFn})-(\ref{LegFenchelConjugate}), it follows that $h_{\mathcal{K}}(\bm{y})$ is the Legendre-Fenchel conjugate of the indicator function
\[\mathbf{1}_{\mathcal{K}}(\bm{x}) := \begin{cases}0 & \text{if}\quad \bm{x}\in\mathcal{K},\\
+\infty &\text{otherwise.}\end{cases}\]
Since the indicator function of a convex set is a convex function, the biconjugate $\mathbf{1}_{\mathcal{K}}^{**}(\cdot) = h_{\mathcal{K}}^{*}(\cdot) = \mathbf{1}_{\mathcal{K}}(\cdot)$. This will be useful in Section \ref{sec:Taxonomy}.

To proceed further, we introduce some notations. Since $\mathcal{U}$ is compact, let
\begin{align}
\alpha_{j} := \underset{\bm{u}\in\mathcal{U}}{\min} \; u_{j}, \quad \beta_{j} := \underset{\bm{u}\in\mathcal{U}}{\max} \; u_{j}, \quad j\in[m],    
\label{defalphabeta}    
\end{align}
that is, $\alpha_{j}$ and $\beta_{j}$ are the component-wise minimum and maximum, respectively, of the input vector. Furthermore, let 
\begin{align}
\mu_{j}:=\frac{\beta_{j}-\alpha_{j}}{2}, \quad \nu_{j}:=\frac{\beta_{j}+\alpha_{j}}{2},
\label{defmujnuj}	
\end{align}
and introduce
\begin{align}
\!\!\bm{\xi}(s):=\!\begin{pmatrix}
\mu_1\bm{\xi}_1(s)\\
\vdots\\
\mu_m\bm{\xi}_m(s)
\end{pmatrix},\;
\bm{\xi}_{j}(s) := \!\begin{pmatrix}
s^{r_j-1}/(r_j-1)! \\
s^{r_j-2}/(r_j-2)! \\
  \vdots\\
  s\\
   1
\end{pmatrix},  
\label{xiVector}
\end{align}
for $j\in[m]$. Also, let 
\begin{align}
\bm{\zeta}(t_0,t) = \begin{pmatrix}
 \mu_{1}\bm{\zeta}_{1}(t_0,t)\\
  \mu_{2}\bm{\zeta}_{2}(t_0,t)\\
  \vdots\\
   \mu_{m}\bm{\zeta}_{m}(t_0,t)	
 \end{pmatrix}, \quad \bm{\zeta}_{j}(t_0,t):=\int_{t_0}^{t}\!\bm{\xi}_{j}(s)\:\differential s\in\mathbb{R}^{r_{j}},
\label{DefzetaDoubleArgument}	
\end{align}
for $j\in[m]$. When $t_{0}=0$, we simplify the notations as
\begin{align}
\!\!\!\!\!\!\bm{\zeta}(t):=\bm{\zeta}(0,t),\; \bm{\zeta}_{j}(t):=\bm{\zeta}_{j}(0,t) \;\text{for all}\; j\in[m].
\label{DefzetaSingleArgument}	
\end{align}
Using \eqref{xiVector} and following \eqref {DefSptFn}, we deduce Proposition \ref{proSptFn} stated next (proof in Appendix \ref{AppendixProofPro:proSptFn}).  
\begin{proposition}\label{proSptFn}
(\textbf{Support function for compact $\mathcal{U}$}) For compact convex $\mathcal{X}_{0}\subset\mathbb{R}^{d}$, and compact $\mathcal{U}\subset\mathbb{R}^{m}$, the support function of the reach set (\ref{SetValuedIntegral}) is
\begin{align}
&h_{\mathcal{R}\left(\mathcal{X}_{0},t\right)}\left(\bm{y}\right) = \underset{\bm{x}_{0}\in\mathcal{X}_{0}}{\sup} 
\sum_{j=1}^{m}~ \langle\bm{y}_{j},\exp\left(t\bm{A}_j\right)\bm{x}_{j0}\rangle\nonumber\\
&+ \int^t_0 \underset{\bm{u}\in{\rm{closure}}\left({\rm{conv}}\left(\mathcal{U}\right)\right)}{\sup}\sum_{j=1}^{m}~\{\langle\bm{y}_{j},\bm{\xi}_{j}(s)\rangle\: u_{j}\} \:\differential{s},
\label{SptFnIntegratorpGen}	
\end{align}	
where ${\rm{conv}}(\cdot)$ denotes the convex hull.
\end{proposition} 
%\underset{\substack{{\bm{x}_{0}\in\mathcal{X}_{0}} \\ {\bm{u}\in\mathcal{U}} }}{\sup}  

\subsubsection{Polar dual}\label{subsubsec:PolarDual}
The \emph{polar dual} $\mathcal{K}^{\circ}$ of any non-empty set $\mathcal{K}\subset\mathbb{R}^{d}$ is given by
\begin{align}
\mathcal{K}^{\circ} := \{\bm{y}\in\mathbb{R}^{d}\mid \langle\bm{y},\bm{x}\rangle \leq 1\quad\text{for all}\;\bm{x}\in\mathcal{K}\}.
\label{DefPolarDual}	
\end{align}
From this definition, it is immediate that $\mathcal{K}^{\circ}$ contains the origin, and is a closed convex set. The bipolar $\left(\mathcal{K}^{\circ}\right)^{\circ} = {\rm{closure}}\left({\rm{conv}}\left(\mathcal{K}\cup\{\bm{0}\}\right)\right)$. Thus, if $\mathcal{K}$ is compact convex and contains the origin, then we have the involution $\left(\mathcal{K}^{\circ}\right)^{\circ} = \mathcal{K}$.
From (\ref{DefSptFn}) and (\ref{DefPolarDual}), notice that $\mathcal{K}^{\circ}$ is the \emph{unit support function ball}, i.e., $\mathcal{K}^{\circ}=\{\bm{y}\in\mathbb{R}^{d}\mid h_{\mathcal{K}}(\bm{y})\leq 1\}$. In Sec. \ref{subsec:DualOfReachSet}, we will mention some properties of the polar dual of the integrator reach set.

\subsubsection{Vector measure}\label{subsubsec:VectorMeasure}
Let $\mathcal{F}$ be a $\sigma$-field of the subsets of a set. A countably additive mapping $\widetilde{\bm{\mu}}:\mathcal{F}\mapsto\mathbb{R}^{d}$ is termed a \emph{vector measure}. Here, ``countably additive" means that for any sequence $\{\Omega_{i}\}_{i=1}^{\infty}$ of disjoint sets in $\mathcal{F}$ such that their union is in $\mathcal{F}$, we have $\widetilde{\bm{\mu}}\left(\cup_{i=1}^{\infty}\Omega_{i}\right)=\sum_{i=1}^{\infty}\widetilde{\bm{\mu}}\left(\Omega_{i}\right) < \infty$. Some of the early investigations of vector measures were due to Liapounoff \cite{liapounoff1940fonctions} and Halmos \cite{halmos1948range}; relatively recent references are \cite{diestel1974vector,dinculeanu2014vector}.

\subsubsection{Zonotope}\label{subsubsec:Zonotope}
A \emph{zonotope} $\mathcal{Z}\subset\mathbb{R}^{d}$ is a finite Minkowski sum of closed line segments or intervals $\{I_{i}\}_{i=1}^{n}$ where these intervals are imbedded in the ambient Euclidean space $\mathbb{R}^{d}$. Explicitly, {{for some positive integer $n$, we write}}
\begin{align*}
\mathcal{Z} &:= I_{1} \dotplus \hdots\dotplus I_{n} \\
&= \bigg\{\bm{x}\in\mathbb{R}^{d}\mid \bm{x} = \sum_{i=1}^{n}\bm{x}_{i}, \quad \bm{x}_{i}\in I_{i}, \quad i\in[n] \bigg\}.	
\end{align*}
Thus, a zonotope is the range of an \emph{atomic} vector measure. Alternatively, a zonotope can be viewed as the affine image of the unit cube. A compact convex polytope is a zonotope if and only if all its two dimensional faces are centrally symmetric \cite[p. 182]{schneider2014convex}. For instance, the cross polytope $\{\bm{x}\in\mathbb{R}^{d}\mid\|\bm{x}\|_{1}\leq 1\}$, is not a zonotope. Standard references on zonotope include \cite{mcmullen1971on,shephard1974combinatorial}, \cite[Ch. 2.7]{coxeter1973regular}. 

The set of zonotopes is closed under affine image and Minkowski sum, but not under intersection. In the systems-control literature, a significant body of work exists on computationally efficient over-approximation of reach sets via zonotopes \cite{girard2005reachability,girard2008zonotope,althoff2010computing} and its variants such as zonotope bundles \cite{althoff2011zonotope}, constrained zonotopes \cite{scott2016constrained}, complex zonotopes \cite{adimoolam2016using}, and polynomial zonotopes \cite{althoff2013reachability,kochdumper2020sparse}.

\subsubsection{Variety and ideal}\label{subsubsec:VarietyIdeal}
Let $p_1,\hdots,p_n\in\mathbb{R}[x_1,\hdots,x_d]$, the vector space of real-valued $d$-variate polynomials. The (affine) \emph{variety} $V_{\mathbb{R}[x_1,\hdots,x_d]}(p_1,\hdots,p_n)$ is the set of all solutions of the system $p_1(x_1,x_2,\hdots,x_d)=\hdots = p_n(x_1,x_2,\hdots,x_d)=0$. Given $p_1,\hdots,p_n\in\mathbb{R}[x_1,\hdots,x_d]$, the set 
\[I:=\bigg\{\sum_{i=1}^{n}\alpha_i p_i \mid \alpha_{1},\hdots,\alpha_{n}\in\mathbb{R}[x_1,\hdots,x_d]\bigg\} \]
is called the \emph{ideal} generated by $p_1,\hdots,p_n$. We write this symbolically as $I=\langle\langle p_1,\hdots,p_n\rangle\rangle$. Roughly speaking, $\langle\langle p_1,\hdots,p_n\rangle\rangle$ is the set of all polynomial consequences of the given system of $n$ polynomial equations in $d$ indeterminates. We refer the readers to \cite[Ch. 1]{cox2013ideals} for detailed exposition of these concepts.

%%%%%%%%%%%%%%%%%%%%%%%%%%%%%%%%%%%%%%%%%%%%%%%%%%%%%%%%%%%%%%%%%%%%%%%%%%%%%%%%%%%%%%%%%%%%%%%%%%%%%%%%%%%%%%%%%%%%%%%%%%%%%%%%%%%%%%%%%%%%%%%%%%%%%%%%%%%

\section{Box-valued Input Uncertainty}\label{sec:BoxValued}
In the remaining of this paper, we characterize the exact reach set \eqref{DefReachSet} when input set $\mathcal{U}\subset\mathbb{R}^{m}$ is box-valued, and remark on the quality of approximation for the same when $\mathcal{U}$ is arbitrary compact.

%From (\ref{blkdiagAB}), the system matrices are block diagonal. 
When $\mathcal{U}\subset\mathbb{R}^{m}$ is box-valued, denote the reach set \eqref{DefReachSet} as $\mathcal{R}^{\Box}$, i.e., with a box superscript\footnote{For the single input ($m=1$) case, we drop the box superscript.}. In this case, each of the $m$ single input integrator dynamics with $r_{j}$ dimensional state subvectors for $j\in[m]$, are decoupled from each other. Then $\mathcal{R}^{\Box}\left(\mathcal{X}_{0},t\right)\subset\mathbb{R}^{d}$ is the Cartesian product of these single input integrator reach sets: $\mathcal{R}_{j}\left(\mathcal{X}_{0},t\right) \subset \mathbb{R}^{r_{j}}$ for $j\in[m]$, i.e.,
\begin{align}
\mathcal{R}^{\Box}=\mathcal{R}_{1}\times\mathcal{R}_{2}\times\hdots\times\mathcal{R}_{m}.
\label{CartesianProduct}	
\end{align}
In what follows, we will sometimes exploit that (\ref{CartesianProduct}) may also be written as\footnote{{{In general, the Minkowski sum of a given collection of compact convex sets is not equal to their Cartesian product. However, the ``factor sets" in \eqref{CartesianProduct} belong to disjoint mutually orthogonal $r_{j}$ dimensional subspaces, $j=1,\hdots,m$, which allows writing this Cartesian product as a Minkowski sum.}}} a Minkowski sum $\mathcal{R}_{1} \dotplus \hdots \dotplus \mathcal{R}_{m}$. Notice that the decoupled dynamics also allows us to write a {{Minkowski}} sum decomposition for the set of initial conditions 
\[\mathcal{X}_{0} = \mathcal{X}_{10} \dotplus \hdots \dotplus \mathcal{X}_{m0},\] and accordingly, the initial condition subvectors $\bm{x}_{j0}\in\mathcal{X}_{j0}\subset\mathbb{R}^{r_j}$ for $j\in[m]$. Thus $\bm{x}_{0}=(\bm{x}_{10},\hdots,\bm{x}_{m0})^{\top}$.

Since the support function of the Minkowski sum is equal to the sum of the support functions, we have
\begin{align}
&h_{\mathcal{R}^{\Box}\left(\mathcal{X}_{0},t\right)}(\bm{y}) = \sum_{j=1}^{m} h_{\mathcal{R}_{j}\left(\mathcal{X}_{j0},t\right)}(\bm{y}_{j}).
\label{SptFn}    
\end{align}
This leads to the following result (proof in Appendix \ref{AppendixProofThm:SptFn}) which will come in handy in the ensuing development.
\begin{theorem}\label{ThmSptFn}
(\textbf{Support function for box-valued $\mathcal{U}$}) For compact convex $\mathcal{X}_{0}\subset\mathbb{R}^{d}$, and box-valued input uncertainty set given by
\begin{align}
\label{BoxInputSet} 
&\mathcal{U}:=\left[\alpha_1,\beta_1\right]\times \left[\alpha_2,\beta_2\right]\times \hdots \times \left[\alpha_m,\beta_m\right]\subset\mathbb{R}^{m}\:,
\end{align}
 the support function of the reach set (\ref{SetValuedIntegral}) is
\begin{align}
\!\!&h_{\mathcal{R}^{\Box}\left(\mathcal{X}_{0},t\right)}\left(\bm{y}\right) =  
\sum_{j=1}^{m}\bigg\{\underset{\bm{x}_{j0}\in\mathcal{X}_{j0}}{\sup}\langle\bm{y}_{j},\exp\left(t\bm{A}\right)\bm{x}_{j0}\rangle\nonumber\\ 
&\qquad\qquad\quad+ \nu_j\langle\bm{y}_{j},\bm{\zeta}_{j}(t)\rangle + \mu_{j}\int_{0}^{t}\!\lvert\langle\bm{y}_{j},\bm{\xi}_{j}(s)\rangle\rvert\:\differential s\bigg\}.
\label{SptFnIntegratorFinal}	
\end{align}	
\end{theorem}
The formula \eqref{SptFnIntegratorFinal} upper bounds \eqref{SptFnIntegratorpGen} resulting from the same
initial condition and arbitrary compact $\mathcal{U}\subset\mathbb{R}^{m}$ with $\{\alpha_j,\beta_j\}_{j=1}^{m}$ related to $\mathcal{U}$ via \eqref{defalphabeta}. Thus, from \eqref{SptFnInequality}, the reach set $\mathcal{R}^{\Box}$ with box-valued input uncertainty will over-approximate the reach set $\mathcal{R}$ associated with arbitrary compact $\mathcal{U}$, at any given $t>0$, provided $\{\alpha_j,\beta_j\}_{j=1}^{m}$ are defined as \eqref{defalphabeta}.

When $\mathcal{U}$ is compact but not box-valued, then we can quantify the quality of the aforesaid over-approximation in terms of the two-sided Hausdorff distance metric $\dist$
%\begin{align}
%&\dist:=\max \left\{\sup _{\boldsymbol{x}(t) \in \mathcal{X}(t)} \inf _{\widehat{\boldsymbol{x}}(t) \in \widehat{\mathcal{E}}(t)}\|\boldsymbol{x}(t)-\widehat{\boldsymbol{x}}(t)\|_{2}\right. \text {, }\nonumber\\
%&\left.\sup _{\widehat{\boldsymbol{x}}(t) \in \widehat{\mathcal{E}}(t)} \inf _{\boldsymbol{x}(t) \in \mathcal{X}(t)}\|\boldsymbol{x}(t)-\widehat{\boldsymbol{x}}(t)\|_{2}\right\}.
%\label{defHausdorff}
%\end{align}
between the convex compact sets $\mathcal{R}^{\Box},\mathcal{R}\subset\mathbb{R}^{d}$, expressible \cite[Thm. 1.8.11]{schneider2014convex} in terms of their support functions $h_{\mathcal{R}^{\Box}}(\cdot),h_{\mathcal{R}}(\cdot)$ as 
\begin{align}
\dist\left(\mathcal{R}^{\Box},\mathcal{R}\right) = \underset{\|\bm{y}\|_{2} = 1}{\sup}\quad \big\vert h_{\mathcal{R}^{\Box}}(\bm{y})- h_{\mathcal{R}}(\bm{y})\big\vert.
\label{HausdorffDist}	
\end{align}
Thanks to \eqref{SptFnInequality}, the absolute value in \eqref{HausdorffDist} can be dispensed since $\mathcal{R}\subseteq \mathcal{R}^{\Box}$ with set equality if $\mathcal{U}$ is box, in which case $h_{\mathcal{R}^{\Box}}(\cdot) = h_{\mathcal{R}}(\cdot)$ and $\dist = 0$.

It is known that \cite[Prop. 6.1]{yong1999stochastic} the set $\mathcal{R}\left(\mathcal{X}_0,t\right)$ resulting from a linear time invariant dynamics such as \eqref{IntegratorDyn}-\eqref{defAB} remains invariant under the closure of convexification of the input set $\mathcal{U}$. Therefore, it is possible that $\mathcal{R}= \mathcal{R}^{\Box}$ and $\dist = 0$ even when the compact set $\mathcal{U}$ is nonconvex. For instance, the reach set $\mathcal{R}\left(\mathcal{X}_0,t\right)$ resulting from some compact convex $\mathcal{X}_0\subset\mathbb{R}^{d}$ and dynamics \eqref{IntegratorDyn}-\eqref{defAB} with the nonconvex input uncertainty set $\{-1,1\}^{m}$, is identical to $\mathcal{R}^{\Box}\left(\mathcal{X}_0,t\right)$ resulting from the same $\mathcal{X}_0$, same dynamics, and the box-valued input uncertainty set \eqref{BoxInputSet} with $\alpha_j=-1, \beta_{j}=1$ for all $j\in[m]$.

Likewise, for the same compact convex $\mathcal{X}_0\subset\mathbb{R}^{d}$, the reach set $\mathcal{R}\left(\mathcal{X}_0,t\right)$ resulting from \eqref{IntegratorDyn}-\eqref{defAB} with the nonconvex input set $\{\bm{u}\in\mathbb{R}^{m}\mid \|\bm{u}\|_{p}\leq 1\}$, $0 < p < 1$, is the same as that resulting from the cross-polytope $\{\bm{u}\in\mathbb{R}^{m}\mid \|\bm{u}\|_{1}\leq 1\}$. More generally, for $0<p<\infty$, suppose $\mathcal{R}_{p}\left(\mathcal{X}_0,t\right)$ results from the unit $p$ norm ball input uncertainty set $\{\bm{u}\in\mathbb{R}^{m}\mid \|\bm{u}\|_{p}\leq 1\}$. Let $\bm{M}^{\top}(\tau):=\exp\left(\tau\bm{A}\right)\bm{B} = \blkdiag\left(\bm{\xi}_{1},\hdots,\bm{\xi}_{m}\right)$. If $\mathcal{R}^{\Box}\left(\mathcal{X}_0,t\right)$ results from the same $\mathcal{X}_0$, same dynamics, and input uncertainty set \eqref{BoxInputSet} with $\alpha_j=-1, \beta_{j}=1$ for all $j\in[m]$, then using \cite[Thm. 1]{haddad2022certifying}, \eqref{HausdorffDist} simplifies to 
\begin{align}
\!\!\!\dist\!\left(\!\mathcal{R}^{\Box},\mathcal{R}_{p}\!\right)\! = \!\!\underset{\|\bm{y}\|_{2} = 1}{\sup} \int_{0}^{t}\!\!\!\left(\|\bm{M}(\tau)\bm{y}\|_{1}\!-\!\|\bm{M}(\tau)\bm{y}\|_{q}\right)\differential\tau  
\label{HausdorffIntegral}    	
\end{align} 
where $q$ is the H\"{o}lder conjugate of $\max\{1,p\}$, i.e., $\frac{1}{\max\{1,p\}}+\frac{1}{q}=1$, and $1<q\leq \infty$. In this case, the positive value \eqref{HausdorffIntegral} quantifies the quality of strict over-approximation $\mathcal{R}_{p}\subset\mathcal{R}^{\Box}$ for $0<p<\infty$. The objective in \eqref{HausdorffIntegral} being positive homogeneous, admits lossless constraint convexification $\|\bm{y}\|_{2} \leq 1$, and the corresponding maximal value\footnote{As such, \eqref{HausdorffIntegral} has a difference of convex objective, and by the Weierstrass extreme value theorem, the maximum is achieved.} for moderate dimensions $d$, can be found by direct numerical search.

%%%%%%%%%%%%%%%%%%%%%%%%%%%%%%%%%%%%%%%%%%%%%%%%%%%%%%%%%%%%%%%%%%%%%%%%%%%%%%%%%%%%%%%%%%%%%%%%%%%%%%%%%%%%%%%%%%%%%%%%%%%%%%%%%%%%%%%%%%%%%%%%%%%%%%%%%%%

\section{Taxonomy and Boundary}\label{sec:Taxonomy}
For $\mathcal{X}_{0}\subset\mathbb{R}^{d}$ compact convex, it is well-known \cite[Sec. 2]{varaiya2000reach} that the reach set $\mathcal{R}$ given by \eqref{DefReachSet} is compact convex for all $t>0$ provided $\mathcal{U}$ is compact. However, it is not immediate what kind of convex set $\mathcal{R}$ is, even for singleton $\mathcal{X}_0 \equiv \{\bm{x}_{0}\}$.

In this Section, we examine the question ``what type of compact convex set $\mathcal{R}^{\Box}\left(\{\bm{x}_{0}\},t\right)$ is" when $\mathcal{U}$ is box-valued uncertainty set of the form \eqref{BoxInputSet}. In the same setting, we also derive the equations for the boundary $\partial\mathcal{R}^{\Box}\left(\{\bm{x}_{0}\},t\right)$. 

Notice that for non-singleton $\mathcal{X}_{0}$, the taxonomy question is not well-posed since the classification then will depend on $\mathcal{X}_{0}$. Also, setting $\mathcal{X}_0 \equiv \{\bm{x}_{0}\}$ in (\ref{SetValuedIntegral}), it is apparent that $\mathcal{R}\left(\{\bm{x}_{0}\},t\right)$ is a translation of the set-valued integral in (\ref{SetValuedIntegral}). Thus, classifying $\mathcal{R}\left(\{\bm{x}_{0}\},t\right)$ amounts to classifying the second summand in (\ref{SetValuedIntegral}).

\subsection{$\mathcal{R}^{\Box}\left(\{\bm{x}_{0}\},t\right)$ is a Zonoid}\label{subsec:zonoid}
A \emph{zonoid} is a compact convex set that is defined as the range of an \emph{atom free} vector measure (see Sec. \ref{subsubsec:VectorMeasure}). Affine image of a zonoid is a zonoid. Minkowski sum of zonoids is also a zonoid. We refer the readers to \cite{bolker1969class,schneider1983zonoids,goodey1993zonoids},\cite[Sec. I]{bourgain1989approximation} for more details on the properties of a zonoid. By slight abuse of nomenclature, in this paper we use the term zonoid up to translation, i.e., we refer to the translation of zonoids as zonoids (instead of using another term such as ``zonoidal translates").

Let us mention a few examples. Any compact convex symmetric set in $\mathbb{R}^{2}$ is a zonoid. In dimensions three or more, all $\ell_{p}$ norm balls for $p\geq 2$ are zonoids.

An alternative way to think about the zonoid is to view it as the limiting set (convergence with respect to the two-sided Hausdorff distance, see e.g., \cite[Appendix B]{haddad2020convex}) of the Minkowski sum of line segments, i.e., the limit of a sequence of \emph{zonotopes} \cite{bolker1969class, mcmullen1971on, shephard1974combinatorial}. {{Formally, given a Hausdorff convergent sequence of zonotopes $\!\{\mathcal{Z}_{j}\}$, the zonoid $\mathcal{Z}_{\infty}$ is
$$\mathcal{Z}_{\infty}:= \lim_{j\rightarrow\infty}\mathcal{Z}_{j}, \;\text{where}\; \mathcal{Z}_{j}  := \sum_{i=1}^{n(j)} \left[\bm{a}_{ij},\bm{b}_{ij}\right], \quad\bm{a}_{ij},\bm{b}_{ij}\in\mathbb{R}^{d},$$
for some $\bm{a}_{ij} \leq \bm{b}_{ij}$ (element-wise vector inequality), and a suitable mapping $n:\mathbb{Z}_{+}\mapsto\mathbb{Z}_{+}$.}} Our analysis will make use of this viewpoint in Sec. \ref{subsec:Volume}. Our main result in this subsection is the following.

\begin{theorem}\label{Thm:zonoid}
The reach set $\mathcal{R}^{\Box}$ given by \eqref{DefReachSet} with $\mathcal{X}_{0}\equiv\{\bm{x}_{0}\}$ and $\mathcal{U}$ given by \eqref{BoxInputSet}, is a zonoid.	
\end{theorem}

To appreciate Theorem \ref{Thm:zonoid} via the limiting viewpoint mentioned before, let us write
%\begin{align}
%\mathcal{R}\left(\{\bm{x}_{0}\},t\right) = \underbrace{\exp(t\bm{A})\bm{x}_{0}\vphantom{\displaystyle\sum_{j=1}^{m}\displaystyle\lim_{n\rightarrow\infty} \displaystyle\sum_{i=0}^{n}\frac{t}{n}\mu_{i}\bm{\xi}_{j}(t_{i})\left[-1,1\right]}}_{\textup{first term}} \dotplus \underbrace{\displaystyle\sum_{j=1}^{m}\displaystyle\lim_{n\rightarrow\infty} \displaystyle\sum_{i=0}^{n}\frac{t}{n}\mu_{j}\bm{\xi}_{j}(t_{i})\left[-1,1\right]}_{\textup{second term}},
%\label{LimitInterpretationZonoid}	
%\end{align}
\begin{align}
\mathcal{R}^{\Box}\left(\{\bm{x}_{0}\},t\right) = &\underbrace{\exp(t\bm{A})\bm{x}_{0}+\displaystyle\sum_{j=1}^{m}\nu_{j}\displaystyle\bm{\zeta}_j(t)\vphantom{\displaystyle\sum_{j=1}^{m}\displaystyle\lim_{n\rightarrow\infty} \displaystyle\sum_{i=0}^{n}\frac{t}{n}\mu_{i}\bm{\xi}_{j}(t_{i})\left[-1,1\right]}}_{\textup{first term}} \nonumber\\
&\dotplus \underbrace{\displaystyle\sum_{j=1}^{m}\displaystyle\lim_{n\rightarrow\infty} \displaystyle\sum_{i=0}^{n}\frac{t}{n}\mu_{j}\displaystyle\bm{\xi}_{j}(t_{i})\left[-1,1\right]}_{\textup{second term}},
\label{LimitInterpretationZonoid}		
\end{align}
where all summation symbols denote Minkowski sums. The first term in (\ref{LimitInterpretationZonoid}) denotes a translation. In the second term, the outer summation over index $j$ arises by writing the Cartesian product (\ref{CartesianProduct}) as the Minkowski sum $\mathcal{R}_{1} \dotplus \hdots \dotplus \mathcal{R}_{m}$. Furthermore, uniformly discretizing $[0,t]$ into $n$ subintervals $[(i-1)t/n, it/n)$, $i=1,\hdots,n$, we write $\int_{0}^{t}\exp(s\bm{A}_{j})\bm{b}_{j}[-\mu_{j},\mu_{j}]\differential s$ as the limit of the Minkowski sum over index $i$. Geometrically, the innermost summands in the second term denote non-uniformly rotated and scaled line intervals in $\mathbb{R}^{j}$. In other words, the second term in (\ref{LimitInterpretationZonoid}) is a Minkowski sum of $m$ sets, each of these sets being the limit of a sequence of sets $\{\mathcal{Z}_{n}\}$  comprising of zonotopes 
\[\mathcal{Z}_{n} := \displaystyle\sum_{i=0}^{n}\frac{t}{n}\mu_{j}\bm{\xi}_{j}(t_{i})\left[-1,1\right],\]
which are the Minkowski sum of $n+1$ line segments. Since $\lim_{n\rightarrow\infty}\mathcal{Z}_{n}$ is a zonoid, the second term in (\ref{LimitInterpretationZonoid}) is a Minkowski sum of $m$ zonoids, and is therefore a zonoid \cite[Thm. 1.5]{bolker1969class}. The entire right hand side of (\ref{LimitInterpretationZonoid}), then, is translation of a zonoid, and hence a zonoid.

\begin{remark}
If $\mathcal{X}_{0}\subset\mathbb{R}^{d}$ is not singleton, but instead a zonoid, then $\mathcal{R}^{\Box}\left(\mathcal{X}_{0},t\right)$ is still a (translated) zonoid. To see this, notice from (\ref{SetValuedIntegral}) and (\ref{SptFnIntegratorFinal}) that
\begin{align}
\mathcal{R}^{\Box}\left(\mathcal{X}_{0},t\right) = \exp(t\bm{A})\mathcal{X}_{0} \dotplus \mathcal{R}^{\Box}\left(\{\bm{0}\},t\right),
\label{ReachSetFromOrginPlusTraslation}	
\end{align}
and that $\exp(t\bm{A})\mathcal{X}_{0}$, being linear image of a zonoid, is a zonoid \cite[Lemma 1.4]{bolker1969class}. Thus, (\ref{ReachSetFromOrginPlusTraslation}) being Minkowski sum of zonoids, is a zonoid too \cite[Thm. 1.5]{bolker1969class}, up to translation. 	
\end{remark}

{{In the following, we derive formulae for the boundary (Proposition \ref{Prop:ParametricRepresentationOfBoundaryPoint} and Sec. \ref{subsec:Implicitization}) and volume (Theorem \ref{ThmVolIntegratorReachSet}) of the integrator reach set with $\mathcal{X}_{0} = \{\bm{x}_{0}\}$ (singleton set). From \eqref{ReachSetFromOrginPlusTraslation}, it is clear that one cannot expect similar closed form formulae for arbitrary compact (or even arbitrary compact convex) $\mathcal{X}_{0}$. In this sense, our closed form formulae are as general as one might hope for. For a specific non-singleton $\mathcal{X}_{0}$, one can use these formulae to first derive the boundary (resp. volume) of $\mathcal{R}^{\Box}\left(\{\bm{0}\},t\right)$, and then use \eqref{ReachSetFromOrginPlusTraslation} to get \emph{numerical estimates} for the boundary (resp. volume) of $\mathcal{R}^{\Box}\left(\mathcal{X}_{0},t\right)$ (cf. Remark \ref{remark:VolNonsingletonX0}).}}

\subsection{$\mathcal{R}^{\Box}\left(\{\bm{x}_{0}\},t\right)$ is Semialgebraic}\label{subsec:semialgebraic}
A set in $\mathbb{R}^{d}$ is called \emph{basic semialgebraic} if it can be written as a finite conjunction of polynomial inequalities and equalities, the polynomials being in $\mathbb{R}\left[x_{1},\hdots,x_{d}\right]$. Finite union of basic semialgebraic sets is called a \emph{semialgebraic set}. A semialgebraic set need not be basic semialgebriac; see e.g., \cite[Example 2.2]{vinzant2020geometry}.

Semialgebraic sets are closed under finitely many unions and intersections, complement, topological closure, polynomial mapping including projection \cite{tarski1998decision,seidenberg1954new}, and Cartesian product. For details on semialgebraic sets, we refer the readers to \cite[Ch. 2]{bochnak2013real}; see \cite[Appendix A.4.4]{blekherman2012semidefinite} for a short summary.

%{\red{In the following, we show that there exist two polynomials $p_{\text{upper}},p_{\text{lower}}\in\mathbb{R}\left[x_{1},\hdots,x_{d}\right]$ such that the compact convex set $\mathcal{R}\left(\{\bm{x}_{0}\},t\right)$ can be expressed as
%\[\{\bm{x}\in\mathbb{R}^{d}\mid p_{\text{upper}}(\bm{x})\leq 0, \; p_{\text{lower}}(\bm{x})\leq 0\},\]
%i.e., a semialgebraic set with boundary $\partial\mathcal{R}\left(\{\bm{x}_{0}\},t\right) = \{\bm{x}\in\mathbb{R}^{d}\mid p_{\text{upper}}(\bm{x})=0\}\cup\{\bm{x}\in\mathbb{R}^{d}\mid p_{\text{lower}}(\bm{x})=0\}$.}} To this end, in 
In Proposition \ref{Prop:ParametricRepresentationOfBoundaryPoint} below, we derive a parametric representation of $\bm{x}^{\text{bdy}}\in\partial\mathcal{R}^{\Box}\left(\{\bm{x}_{0}\},t\right)$, the boundary of the reach set. Then we use this representation to establish semialgebraicity of $\mathcal{R}^{\Box}\left(\{\bm{x}_{0}\},t\right)$ in Theorem \ref{Thm:semialgebraic} that follows.  

\begin{proposition}\label{Prop:ParametricRepresentationOfBoundaryPoint}
For relative degree vector $\bm{r}=(r_1,\hdots,r_m)^{\top}$, and fixed $\bm{x}_{0}\in\mathbb{R}^{d}$ comprising of subvectors $\bm{x}_{j0}\in\mathbb{R}^{r_{j}}$ where $j\in[m]$, consider the reach set (\ref{SetValuedIntegral}) with singleton $\mathcal{X}_{0}\equiv\{\bm{x}_{0}\}$ and $\mathcal{U}$ given by \eqref{BoxInputSet}. For $j\in[m]$, define $\mu_{1},\hdots,\mu_{m}$ and $\nu_{1},\hdots,\nu_{m}$ as in (\ref{defalphabeta})-(\ref{defmujnuj}). Let the indicator function $\mathbf{1}_{k\leq \ell}:=1$ for $k\leq \ell$, and $:=0$ otherwise.
Then the components of 
\[\bm{x}^{\textup{bdy}}=\begin{pmatrix}
\bm{x}^{\textup{bdy}}_{1}\\
\bm{x}^{\textup{bdy}}_{2}\\
\vdots\\
\bm{x}^{\textup{bdy}}_{m}	
\end{pmatrix}
\in\partial\mathcal{R}^{\Box}\left(\{\bm{x}_{0}\},t\right),\;\bm{x}^{\textup{bdy}}_{j}\in\mathbb{R}^{r_j}, \; j\in[m],\] 
admit {{parametric representation in terms of the parameters $(s_{1}, s_{2}, \hdots, s_{r_{j}-1})$ satisfying $0\leq s_{1} \leq s_{2} \leq \hdots \leq s_{r_{j}-1}\leq t$. This parameterization is given by}}
{\small{\begin{align}
&\bm{x}^{\textup{bdy}}_{j}\left(k\right) =  \sum_{\ell = 1}^{r_{j}}\mathbf{1}_{k\leq \ell}\:\frac{t^{\ell-k}}{(\ell-k)!}\:\bm{x}_{j0}(\ell) +\frac{\nu_{j}\:t^{r_{j}-k+1}}{(r_{j}-k+1)!} \nonumber\\
&\pm \frac{\mu_{j}}{(r_{j}-k+1)!}
\bigg\{(-1)^{r_{j}-1}\:t^{r_{j}-k+1}+2\sum_{q=1}^{r_{j}-1}\!\!(-1)^{q+1}\:s_{q}^{r_{j}-k+1}\bigg\},
\label{ParamRepresentationBoundary}	
\end{align}}}	
where $\bm{x}^{\textup{bdy}}_{j}\left(k\right)$ denotes the $k$th component of the $j$th subvector $\bm{x}^{\textup{bdy}}_{j}$ for $k\in[r_{j}]$.
\end{proposition}
%The proof of the above is deferred to Appendix \ref{AppendixProofProp:ParametricRepresentationOfBoundaryPoint}. 
\noindent The following is a consequence of the $\pm$ appearing in \eqref{ParamRepresentationBoundary}.

\begin{corollary}\label{Corollary:TwoBoundingSurfaces}
The single input integrator reach set $\mathcal{R}_{j}\left(\{\bm{x}_{0}\},t\right) \subset \mathbb{R}^{r_{j}}$ has two bounding surfaces for each $j\in[m]$. In other words, there exist $p_{j}^{\textup{upper}},p_{j}^{\textup{lower}}:\mathbb{R}^{r_{j}}\mapsto\mathbb{R}$ such that
\[\mathcal{R}_{j}\left(\{\bm{x}_{0}\},t\right) = \{\bm{x}\in\mathbb{R}^{r_{j}}\mid p_{j}^{\textup{upper}}(\bm{x})\leq 0, \; p_{j}^{\textup{lower}}(\bm{x})\leq 0\},\]
with boundary $\partial\mathcal{R}_{j}\left(\{\bm{x}_{0}\},t\right) = \{\bm{x}\in\mathbb{R}^{r_{j}}\mid p_{j}^{\textup{upper}}(\bm{x})=0\}\cup\{\bm{x}\in\mathbb{R}^{r_{j}}\mid p_{j}^{\textup{lower}}(\bm{x})=0\}$. 	
\end{corollary}

\begin{figure}[tpb]
        \centering
        \includegraphics[width=0.75\linewidth]{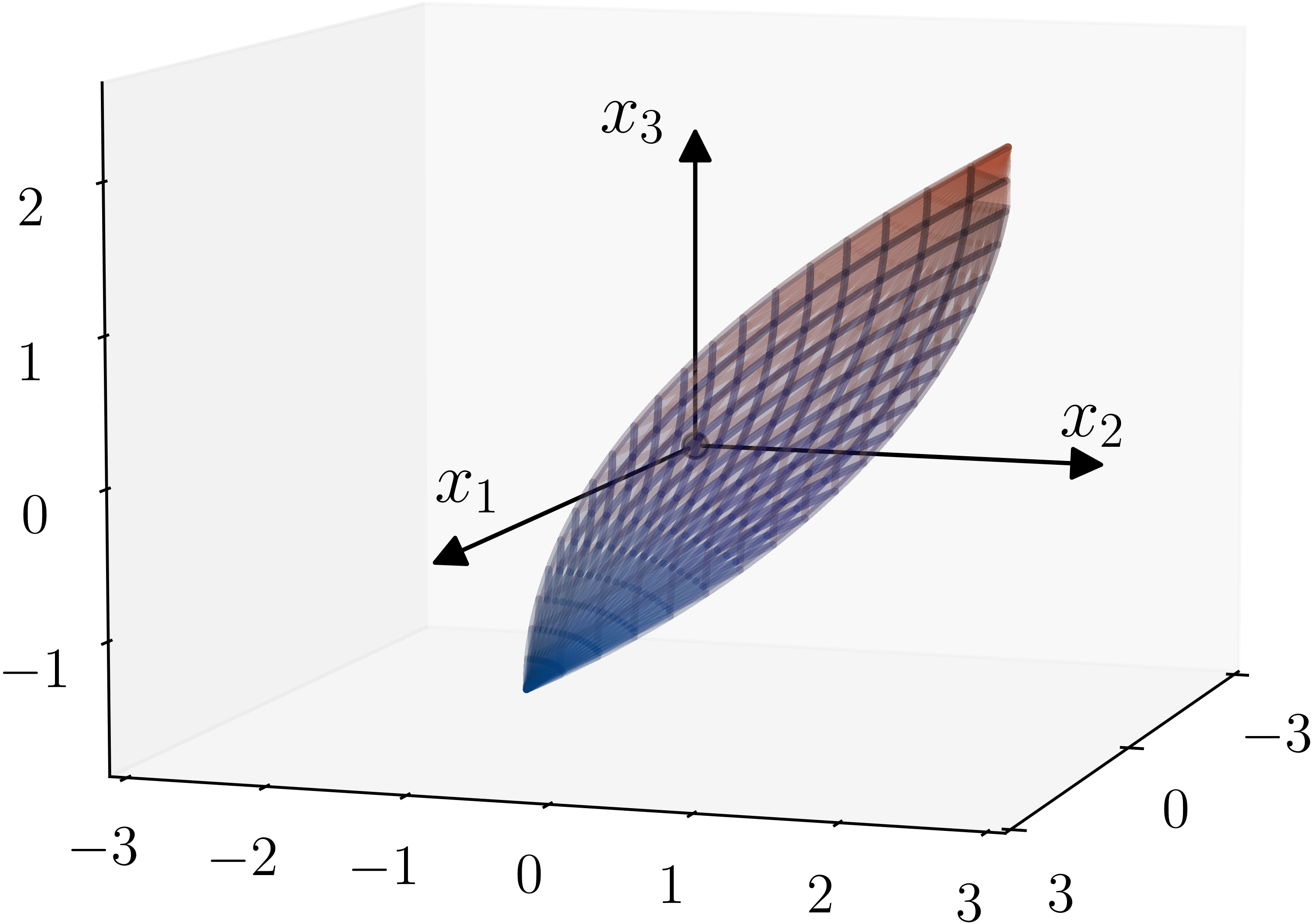}
        \caption{{\small{The ``almond-shaped" integrator reach set $\mathcal{R}(\{\bm{x}_{0}\},t)\subset\mathbb{R}^{3}$ with $d=3$, $m=1$, $\bm{x}_{0} = (0.1,0.2,0.3)^{\top}$, $\mathcal{U}\equiv[\alpha,\beta]=[-1,1]$ at $t=2.1$. The wireframes correspond to the upper and lower surfaces.}}}
\vspace*{-0.2in}
\label{FigSingleInputReachSet3D}
\end{figure}

During the proof of Theorem \ref{Thm:semialgebraic} below, it will turn out that in fact $p_{j}^{\text{upper}},p_{j}^{\text{lower}}\in\mathbb{R}\left[x_{1},\hdots,x_{r_{j}}\right]$ for all $j\in[m]$. In words, $p_{j}^{\text{upper}},p_{j}^{\text{lower}}$ are real algebraic hypersurfaces for all $j\in[m]$. 

Let us exemplify the parameterization (\ref{ParamRepresentationBoundary}) for the case $\bm{r} = (r_{1},r_{2})^{\top} =(2,3)^{\top}$. In this case,
\begin{align}
\!\!\begin{pmatrix}
\bm{x}^{\text{bdy}}_{1}(1)\\
\vspace*{-0.05in}\\
\bm{x}^{\text{bdy}}_{1}(2)	
\end{pmatrix}
 \!\!=\!\!\begin{pmatrix}
 \bm{x}_{10}(1) + t\bm{x}_{10}(2) + \nu_{1}(t^{2}/2) \pm\mu_{1}\left(s_{1}^{2}-t^{2}/2\right)\\
 \vspace*{-0.05in}\\
 \bm{x}_{10}(2) + \nu_{1}t \pm \mu_{1}\left(2s_{1}-t\right)	
 \end{pmatrix},
\label{DoubleIntegratorParametric}	
\end{align}
%and
\begin{align}
\!\!\begin{pmatrix}
\bm{x}^{\text{bdy}}_{2}(1)\\
\vspace*{-0.05in}\\
\bm{x}^{\text{bdy}}_{2}(2)\\
\vspace*{-0.05in}\\
\bm{x}^{\text{bdy}}_{2}(3)	
\end{pmatrix}
 = \begin{pmatrix}
 \bm{x}_{20}(1) + t\bm{x}_{20}(2) + (t^{2}/2)\bm{x}_{20}(3) \\
 + \nu_{2}(t^{3}/6) \pm \mu_{2}\left(t^{3}/6 + 2s_{1}^{3}/6 - 2s_{2}^{3}/6 \right)\\
 \vspace*{-0.05in}\\
 \bm{x}_{20}(2) + t\bm{x}_{20}(3) + \nu_{2}(t^{2}/2) \pm \mu_{2}\left(t^{2}/2 \right.\\
 \left.+ 2s_{1}^{2}/2 - 2s_{2}^{2}/2\right)\\
 \vspace*{-0.05in}\\
  \bm{x}_{20}(3) + \nu_{2}t \pm \mu_{2}\left(t + 2s_{1} - 2s_{2}\right) 
 \end{pmatrix}. \label{TripleIntegratorParametric} 	
\end{align}
In (\ref{DoubleIntegratorParametric}), taking plus (resp. minus) signs in each of component gives the parametric representation of the curve $p_{1}^{\text{upper}} = 0$ (resp. $p_{1}^{\text{lower}}=0$). These curves are as in \cite[Fig. 1(a)]{haddad2020convex}, and their union defines $\partial\mathcal{R}_{1}$. We note that the parameterization (\ref{DoubleIntegratorParametric}) appeared in \cite[p. 111]{kurzhanski2014dynamics}.

Likewise, in (\ref{TripleIntegratorParametric}), taking plus (resp. minus) signs in each of component gives the parametric representation of the surface $p_{2}^{\text{upper}}(\bm{x})=0$ (resp. $p_{2}^{\text{lower}}=0$). The resulting set $\mathcal{R}_{2}$ is the triple integrator reach set, and is shown in Fig. \ref{FigSingleInputReachSet3D}. 
 
Now we come to the main result of this subsection.

\begin{theorem}\label{Thm:semialgebraic}
The reach set $\mathcal{R}^{\Box}$ given by \eqref{DefReachSet} with $\mathcal{X}_{0}\equiv\{\bm{x}_{0}\}$ and $\mathcal{U}$ as in \eqref{BoxInputSet}, is semialgebraic.	
\end{theorem}

Let us illustrate the bounding curves and surfaces for (\ref{DoubleIntegratorParametric}) and (\ref{TripleIntegratorParametric}) respectively, in the implicit form. Eliminating the parameter $s_1$ from (\ref{DoubleIntegratorParametric}) reveals that $p_{1}^{\text{upper}},p_{1}^{\text{lower}}$ are parabolas. In particular,
\begin{align}
&p_{1}^{\text{upper}}(\bm{x}^{\text{bdy}}_{1}(1),\bm{x}^{\text{bdy}}_{1}(2)) = \frac{1}{4}\left(\frac{\bm{x}^{\text{bdy}}_{1}(2) - \bm{x}_{10}(2) -\nu_{1}t}{\mu_1} + t\right)^{\!2}  \nonumber\\
&\quad - \frac{\bm{x}^{\text{bdy}}_{1}(1) - \bm{x}_{10}(1) - t\bm{x}_{10}(2) - \nu_{1}\frac{t^{2}}{2}}{\mu_1} -\frac{t^{2}}{2},
\label{p1Implicit}		
\end{align}
%\begin{subequations}
%\begin{align}
%&p_{1}^{\text{upper}}(\bm{x}^{\text{bdy}}_{1}(1),\bm{x}^{\text{bdy}}_{1}(2)) = \frac{1}{4}\left(\frac{\bm{x}^{\text{bdy}}_{1}(2) - \bm{x}_{10}(2) -\nu_{1}t}{\mu_1} + t\right)^{\!2}  \nonumber\\
%&\quad - \frac{\bm{x}^{\text{bdy}}_{1}(1) - \bm{x}_{10}(1) - t\bm{x}_{10}(2) - \nu_{1}\frac{t^{2}}{2}}{\mu_1} -\frac{t^{2}}{2},\label{p1upper}\\
%&p_{1}^{\text{lower}}(\bm{x}^{\text{bdy}}_{1}(1),\bm{x}^{\text{bdy}}_{1}(2)) = -\frac{1}{4}\!\left(\!-\frac{\bm{x}^{\text{bdy}}_{1}(2) - \bm{x}_{10}(2) -\nu_{1}t}{\mu_1} + t\!\right)^{\!2}\nonumber\\
%&\quad - \frac{\bm{x}^{\text{bdy}}_{1}(1) - \bm{x}_{10}(1) - t\bm{x}_{10}(2) - \nu_{1}\frac{t^{2}}{2}}{\mu_1} +\frac{t^{2}}{2}.\label{p1lower} 	
%\end{align}
%\label{p1Implicit}	
%\end{subequations}
{{and the formula for $p_{1}^{\text{lower}}(\bm{x}^{\text{bdy}}_{1}(1),\bm{x}^{\text{bdy}}_{1}(2))$ follows mutatis mutandis.}} 

Similarly, eliminating the parameters $s_1,s_2$ from (\ref{TripleIntegratorParametric}) reveals that $p_{2}^{\text{upper}},p_{2}^{\text{lower}}$ are quartic polynomials. In particular, %we have
\begin{align}
&p_{2}^{\text{upper}}(\bm{x}^{\text{bdy}}_{2}(1),\bm{x}^{\text{bdy}}_{2}(2),\bm{x}^{\text{bdy}}_{2}(3)) \nonumber\\
&= \frac{1}{16}\left(\frac{\bm{x}^{\text{bdy}}_{2}(3) - \bm{x}_{20}(3) - \nu_{2}t}{\mu_2} - t\right)^{\!4}  \nonumber\\
&\quad + 3\left(\frac{\bm{x}^{\text{bdy}}_{2}(2) - \bm{x}_{20}(2) - t\bm{x}_{20}(3) -\nu_{2}\frac{t^{2}}{2}}{\mu_2} -\frac{t^{2}}{2}\right)^{\!2}\nonumber\\
&\quad-6 \left(\frac{\bm{x}^{\text{bdy}}_{2}(1) - \bm{x}_{20}(1) - t\bm{x}_{20}(2) - \frac{t^{2}}{2}\bm{x}_{20}(3)-\nu_{2}\frac{t^{3}}{6}}{\mu_2} - \frac{t^{3}}{6}\right)\nonumber\\
&\qquad\times\left(\frac{\bm{x}^{\text{bdy}}_{2}(3) - \bm{x}_{20}(3) -\nu_{2}t}{\mu_2} - t\right),
\label{p2Implicit}
\end{align}
{{and again, the formula for $p_{2}^{\text{lower}}(\bm{x}^{\text{bdy}}_{2}(1),\bm{x}^{\text{bdy}}_{2}(2),\bm{x}^{\text{bdy}}_{2}(3))$ follows mutatis mutandis}}.

%\begin{subequations}
%\begin{align}
%&p_{2}^{\text{upper}}(\bm{x}^{\text{bdy}}_{2}(1),\bm{x}^{\text{bdy}}_{2}(2),\bm{x}^{\text{bdy}}_{2}(3)) \nonumber\\
%&= \frac{1}{16}\left(\frac{\bm{x}^{\text{bdy}}_{2}(3) - \bm{x}_{20}(3) - \nu_{2}t}{\mu_2} - t\right)^{\!4}  \nonumber\\
%&\quad + 3\left(\frac{\bm{x}^{\text{bdy}}_{2}(2) - \bm{x}_{20}(2) - t\bm{x}_{20}(3) -\nu_{2}\frac{t^{2}}{2}}{\mu_2} -\frac{t^{2}}{2}\right)^{\!2}\nonumber\\
%&\quad-6 \left(\frac{\bm{x}^{\text{bdy}}_{2}(1) - \bm{x}_{20}(1) - t\bm{x}_{20}(2) - \frac{t^{2}}{2}\bm{x}_{20}(3)-\nu_{2}\frac{t^{3}}{6}}{\mu_2} - \frac{t^{3}}{6}\right)\nonumber\\
%&\qquad\times\left(\frac{\bm{x}^{\text{bdy}}_{2}(3) - \bm{x}_{20}(3) -\nu_{2}t}{\mu_2} - t\right),\label{p2upper}\\
%&p_{2}^{\text{lower}}(\bm{x}^{\text{bdy}}_{2}(1),\bm{x}^{\text{bdy}}_{2}(2),\bm{x}^{\text{bdy}}_{2}(3)) \nonumber\\
%&= -\frac{1}{16}\left(\frac{\bm{x}^{\text{bdy}}_{2}(3) - \bm{x}_{20}(3) - \nu_{2}t}{\mu_2} + t\right)^{\!4}\nonumber\\
%&\quad - 3\left(\frac{\bm{x}^{\text{bdy}}_{2}(2) - \bm{x}_{20}(2) - t\bm{x}_{20}(3) -\nu_{2}\frac{t^{2}}{2}}{\mu_2} +\frac{t^{2}}{2}\right)^{\!2}\nonumber\\
%&\quad + 6 \left(\frac{\bm{x}^{\text{bdy}}_{2}(1) - \bm{x}_{20}(1) - t\bm{x}_{20}(2) - \frac{t^{2}}{2}\bm{x}_{20}(3)-\nu_{2}\frac{t^{3}}{6}}{\mu_2} + \frac{t^{3}}{6}\right)\nonumber\\
%&\qquad\times\left(\frac{\bm{x}^{\text{bdy}}_{2}(3) - \bm{x}_{20}(3) - \nu_{2}t}{\mu_2} + t\right).\label{p2lower} 	
%\end{align}
%\label{p2Implicit}	
%\end{subequations}
A natural question is whether one can generalize the implicitizations as in (\ref{p1Implicit}), (\ref{p2Implicit}) {{to arbitrary state dimensions}}. This is what we address next.

% \begin{figure*}[t]
%         \centering
%         \includegraphics[width=0.8\linewidth]{NonUniqueIntegratorReachSet.pdf}
%         \caption{{\small{{{\emph{Left}: The reach set in $\mathbb{R}^{2}$ at $t=4$ for the double integrator with two inputs, resulting from singleton initial condition $\{\bm{0}\}$. \emph{Right}: An integrator reach set in $\mathbb{R}^{3}$ at $t=4$ for the relative degree $\bm{r}=(2,1)^{\top}$ with two inputs, resulting from singleton initial condition $\{\bm{0}\}$. In both the plots shown above, the solid boundaries were plotted using Proposition 1. The scatter plots were obtained by propagating the states with 4000 randomly sampled input trajectories from the sets $\mathcal{U}_{p}$ given by \eqref{pnormball}. Specifically, we divided the time interval $[0,4]$ into four subintervals $[t_{k-1},t_{k}]$ for $k=1,\hdots,4$, $t_{0}:=0,t_{4}:=4$. We used the \emph{piecewise constant} inputs $\bm{u}(\tau) = \sum_{k=1}^{4}u_{k-1,k}\left(H(\tau-t_{k-1}) - H(\tau-t_{k})\right)$ for $\tau\in[0,4]$, where $H(\cdot)$ denotes the Heaviside step function, and $u_{k-1,k}$ were sampled uniformly at random from the two dimensional $\ell_2$ and $\ell_{\infty}$ unit norm balls (i.e., resampled for each $k=1,\hdots,4$). These numerics support our reasoning in Sec. \ref{subsec:DependenceOnU} that with all other parameters being fixed, the integrator reach sets resulting from the input sets $\mathcal{U}_{2}$ and $\mathcal{U}_{\infty}$ are the same.}}}}}
% \vspace*{-0.2in}
% \label{fig:nonunique}
% \end{figure*}

\subsection{Implicitization of $\partial\mathcal{R}^{\Box}\left(\{\bm{x}_{0}\},t\right)$}\label{subsec:Implicitization}
To derive the implicit equations for the bounding algebraic hypersurfaces $p_{j}^{\text{upper}},p_{j}^{\text{lower}}\in\mathbb{R}\left[x_{1},\hdots,x_{r_{j}}\right]$ for all $j\in[m]$, we need to eliminate the parameters $\left(s_{1},s_{2},\hdots,s_{r_{j}-1}\right)$ from (\ref{ParamRepresentationBoundary}). For this purpose, it is helpful to write (\ref{ParamRepresentationBoundary}) succinctly as
\begin{align}
\rho_{j,k}^{\pm} = \displaystyle\sum_{q=1}^{r_{j}-1}(-1)^{q+1}\:s_{q}^{r_{j}-k+1}, \qquad k\in[r_{j}],
\label{SuccinctParam}	
\end{align}
where
\begin{align}
\rho_{j,k}^{\pm} := &\frac{(r_{j}-k+1)!}{2\mu_{j}}\bigg\{ \bm{x}^{\textup{bdy}}_{j}\left(k\right) -  \sum_{\ell = 1}^{r_{j}}\mathbf{1}_{k\leq \ell}\:\frac{t^{\ell-k}}{(\ell-k)!}\:\bm{x}_{j0}(\ell)\bigg\}\nonumber\\
&-\frac{1}{2}\bigg\{\pm(-1)^{r_{j}-1}\:t^{r_{j}-k+1}+\frac{\nu_{j}}{\mu_{j}}\:t^{r_{j}-k+1} \bigg\}.
\label{defrho}	
\end{align}
To simplify the rather unpleasant notation $\rho_{j,k}^{\pm}$, we will only address the $m=1$ case. In (\ref{SuccinctParam}), this allows us to replace $r_{j}$ by $d$, and to drop the subscript $j$ from the $\rho$'s. This does not invite any loss of generality in terms of implicitization since post derivation, we can replace $d$ by $r_{j}$ to recover the respective $p_{j}$'s.

With slight abuse of notation, we will also drop the superscript $\pm$ from the $\rho$'s in (\ref{SuccinctParam}). Recall that the plus (resp. minus) superscript in the $\rho$'s indicates $p_{j}^{\text{upper}}$ (resp. $p_{j}^{\text{lower}}$). From (\ref{defrho}), it is clear that in either case, the $\rho_{j,k}$ is affine in $\bm{x}^{\textup{bdy}}_{j}\left(k\right)$, which is the $k$th coordinate of the boundary point for the $j$th block. Importantly, for $k\in[r_{j}]$, the quantity $\rho_{j,k}$ does not depend on any other component of the boundary point than the $k$th component. Again, the plus-minus superscripts can be added back post implicitization.

Thus, the notationally simplified version of (\ref{SuccinctParam}) that suffices for implicitization, is       
\begin{align}
\rho_{k} = \displaystyle\sum_{q=1}^{d-1}(-1)^{q+1}\:s_{q}^{d-k+1}, \qquad k=1,\hdots,d,
\label{SimplifiedSuccinctParam}	
\end{align}
which is a system of $d$ homogeneous polynomials in variables $\left(s_{1},s_{2},\hdots,s_{d-1}\right)$. The objective is to derive the implicitized polynomial $\wp(\rho_1,\rho_2,\hdots,\rho_{d})$ associated with (\ref{SimplifiedSuccinctParam}).

When $d=2$, the parameterization \eqref{SimplifiedSuccinctParam} becomes
\begin{align*}
\rho_{1} = s_{1}^{2},\quad \rho_{2} = s_{1},	
\end{align*}
and we get degree 2 implicitized polynomial
\begin{align}
\wp(\rho_1,\rho_2) = \rho_2^2 - \rho_1 = 0.	
\label{pi2d}	
\end{align}
For $k=1,2$, substituting for the $\rho_{1},\rho_{2}$ in (\ref{pi2d}) from (\ref{defrho}) with appropriate plus-minus signs recovers (\ref{p1Implicit}).

When $d=3$, the parameterization \eqref{SimplifiedSuccinctParam} becomes
\begin{align*}
\rho_{1} = s_{1}^{3} - s_{2}^{3},\quad \rho_{2} = s_{1}^{2} - s_{2}^{2},\quad
\rho_{3} = s_{1} - s_{2},	
\end{align*}
elementary algebra gives degree 4 implicitized polynomial 
\begin{align}
\wp(\rho_1,\rho_2,\rho_3) = \rho_{3}^{4}- 4 \rho_{3}\rho_{1} + 3\rho_{2}^{2} = 0.
\label{pi3d}	
\end{align} 
As before, for $k=1,2,3$, substituting for the $\rho_{1},\rho_{2},\rho_{3}$ in (\ref{pi3d}) from (\ref{defrho}) with appropriate plus-minus signs recovers (\ref{p2Implicit}). However, for $d=4$ or higher, it is practically impossible to derive the implicitization via brute force algebra.

A principled way to implicitize (\ref{SimplifiedSuccinctParam}) is due to G. Zaimi \cite{381335}, and starts with defining $\lambda_{k}:=\rho_{d-k+1}$ for $k=1,\hdots,d$. Introduce the sequence $A_k(s_1,s_2,\dots,s_{d-1})$ via the generating function (see e.g., \cite[Ch. 1]{wilf2005generatingfunctionology})
\begin{align}
F(\tau)=\sum_{k\geq 0} A_k \tau^k=\frac{(1-s_1\tau)(1-s_3\tau)\cdots}{(1-s_2\tau)(1-s_4\tau)\cdots}.
\label{GeneratingFn}	
\end{align}
Taking the logarithmic derivative of (\ref{GeneratingFn}), and then using the generating functions $(1-s_{q}\tau)^{-1} = \sum_{k\geq 0}\left(s_{q}\tau\right)^{k}$ for all $q=1,\hdots,d-1$, yields 
\begin{align}
\!\!\frac{F^{\prime}(\tau)}{F(\tau)} = -s_{1}\!\sum_{k\geq 0}\!\left(s_{1}\tau\right)^{k} \!+s_{2}\!\sum_{k\geq 0}\!\left(s_{2}\tau\right)^{k} \!-s_{3}\!\sum_{k\geq 0}\!\left(s_{3}\tau\right)^{k} \!+ \hdots.
\label{LogarithmicDerivative}	
\end{align}
Integrating (\ref{LogarithmicDerivative}) with respect to $\tau$, we obtain
\begin{align}
F(\tau) = \exp\left(-\sum_{k=1}^{d}\frac{\lambda_{k}}{k}\tau^{k}\right).
\label{ExponentialOfPoly}	
\end{align}
Equating (\ref{GeneratingFn}) and (\ref{ExponentialOfPoly}) allows us to compute $A_{k}$ as a degree $k$ polynomial of the $\lambda$'s.

On the other hand, since the generating function (\ref{GeneratingFn}) is a rational function with denominator polynomial of degree $\delta := \lfloor \frac{d-1}{2}\rfloor$, the following Hankel determinant vanishes\footnote{This result goes back to Kronecker \cite{kronecker1881theorie}. See also \cite[p. 5, Lemma III]{salem1983algebraic}.}
\begin{align}
\det[A_{d-2\delta+i+j}]_{i,j=0}^{\delta}=0.
\label{HankelDet}	
\end{align}
Substituting the $A_{k}$'s obtained as degree $k$ polynomials of the $\lambda$'s into \eqref{HankelDet} gives an implicit polynomial in indeterminate $(\lambda_1,\hdots,\lambda_{d})$ of degree $(\delta+1)(d-\delta)$. Finally, reverting back the $\lambda$'s to the $\rho$'s result in the desired implicit polynomial $\wp(\rho_1,\rho_2,\hdots,\rho_{d})$, which is also of degree $(\delta+1)(d-\delta)$.

For instance, when $d=3$, the relation (\ref{HankelDet}) becomes
\begin{align}
\det\left(\begin{bmatrix}
A_{1} & A_{2}\\
A_{2} & A_{3}	
\end{bmatrix}
\right) = 0.
\label{HankelDetFor3d}	
\end{align}
In this case, equating (\ref{GeneratingFn}) and (\ref{ExponentialOfPoly}) gives
$$A_{1}=-\lambda_{1},\; A_{2}=\frac{1}{2}\lambda_{1}^{2}-\frac{1}{2}\lambda_{2}, \; A_{3}=-\frac{1}{6}\lambda_{1}^{3}+\frac{1}{2}\lambda_{1}\lambda_{2}-\frac{1}{3}\lambda_{3}.$$
Substituting these back in (\ref{HankelDetFor3d}) yields the quartic polynomial $\lambda_{1}^{4} + 3\lambda_{2}^{2}-4\lambda_{3}\lambda_{1}=0$, which under the mapping $(\lambda_1,\lambda_2,\lambda_3)\mapsto (\rho_3,\rho_2,\rho_1)$ recovers (\ref{pi3d}), and thus (\ref{p2Implicit}). 

In summary, (\ref{HankelDet}) is the desired implicitization of the bounding hypersurfaces of the single input integrator reach set (up to the change of variables). The Cartesian product of these implicit hypersurfaces gives the implicitization in the multi input case.

\subsection{Dual of $\mathcal{R}\left(\{\bm{x}_{0}\},t\right)$}\label{subsec:DualOfReachSet}
From convex geometry standpoint, it is natural to ask what kind of characterization is possible for the polar dual (see Sec. \ref{subsubsec:PolarDual}) of the integrator reach set $\mathcal{R}$ or $\mathcal{R}^{\Box}$. We know in general that $\mathcal{R}^{\circ}$ will be a closed convex set. Depending on the choice of $\bm{x}_{0},\mathcal{U}$ and $t$, the set $\mathcal{R}\left(\{\bm{x}_{0}\},t\right)$ may not contain the origin, and thus the bipolar 
\[\left(\mathcal{R}\left(\{\bm{x}_{0}\},t\right)\right)^{\circ\circ} = {\rm{closure}}\left({\rm{conv}}\left(\mathcal{R}\left(\{\bm{x}_{0}\},t\right)\cup\{\bm{0}\}\right)\right),\]
that is, we do not have the involution in general.

Furthermore, since $\mathcal{R}^{\Box}\left(\{\bm{x}_{0}\},t\right)$ is semialgebraic from Sec. \ref{subsec:semialgebraic}, so must be its polar dual $\left(\mathcal{R}^{\Box}\left(\{\bm{x}_{0}\},t\right)\right)^{\circ}$; see e.g., \cite[Ch. 5, Sec. 5.2.2]{blekherman2012semidefinite}. 

\begin{figure}[tpb]
\begin{subfigure}{.24\textwidth}
  \centering
  \includegraphics[width=.99\linewidth]{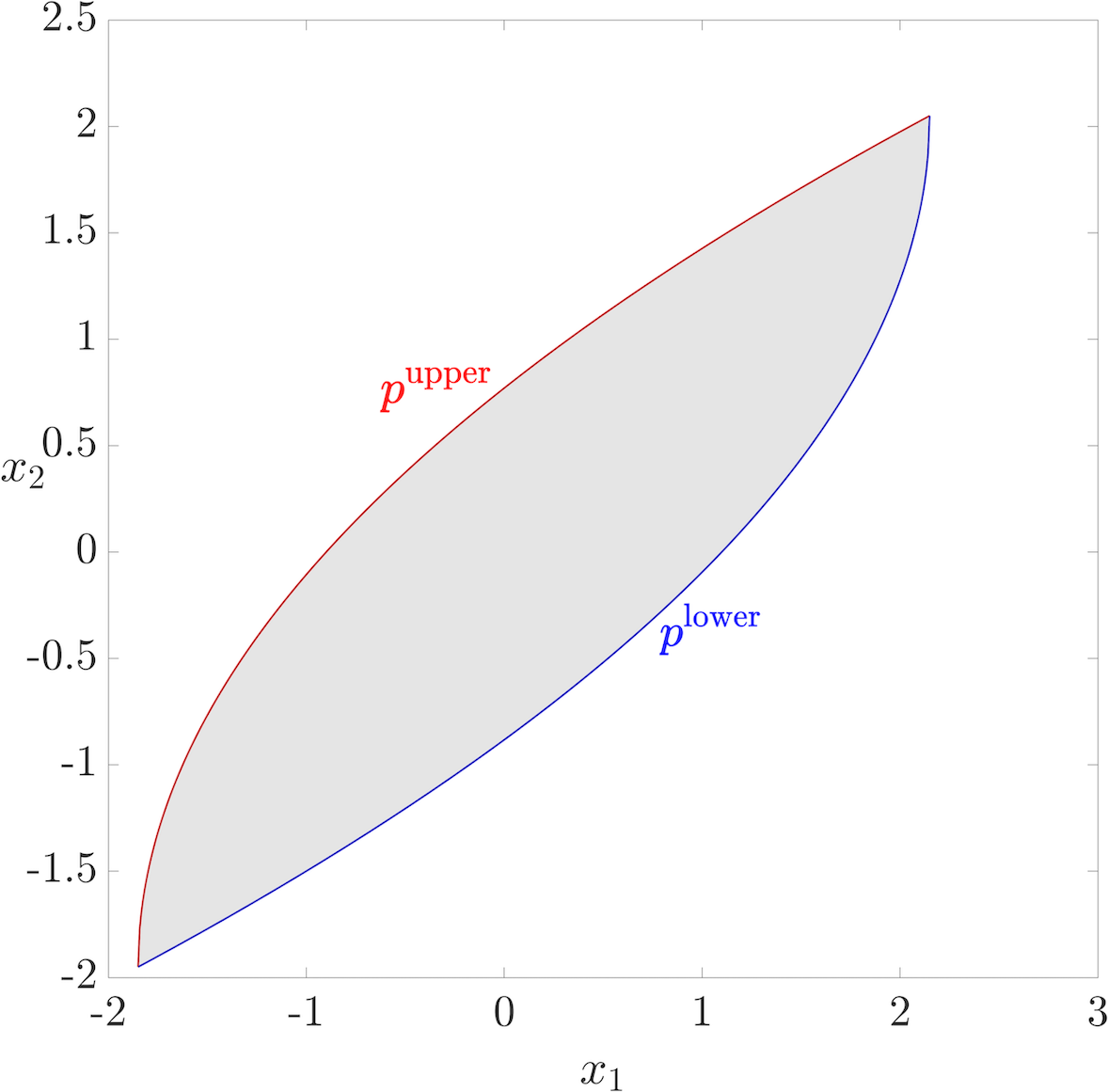}
  \caption{{\small{$\mathcal{R}$ with $\bm{x}_{0}=(0.05,0.05)^{\top}$.}}}
  \label{fig:sfig1}
\end{subfigure}
\begin{subfigure}{.24\textwidth}
  \centering
  \includegraphics[width=.99\linewidth]{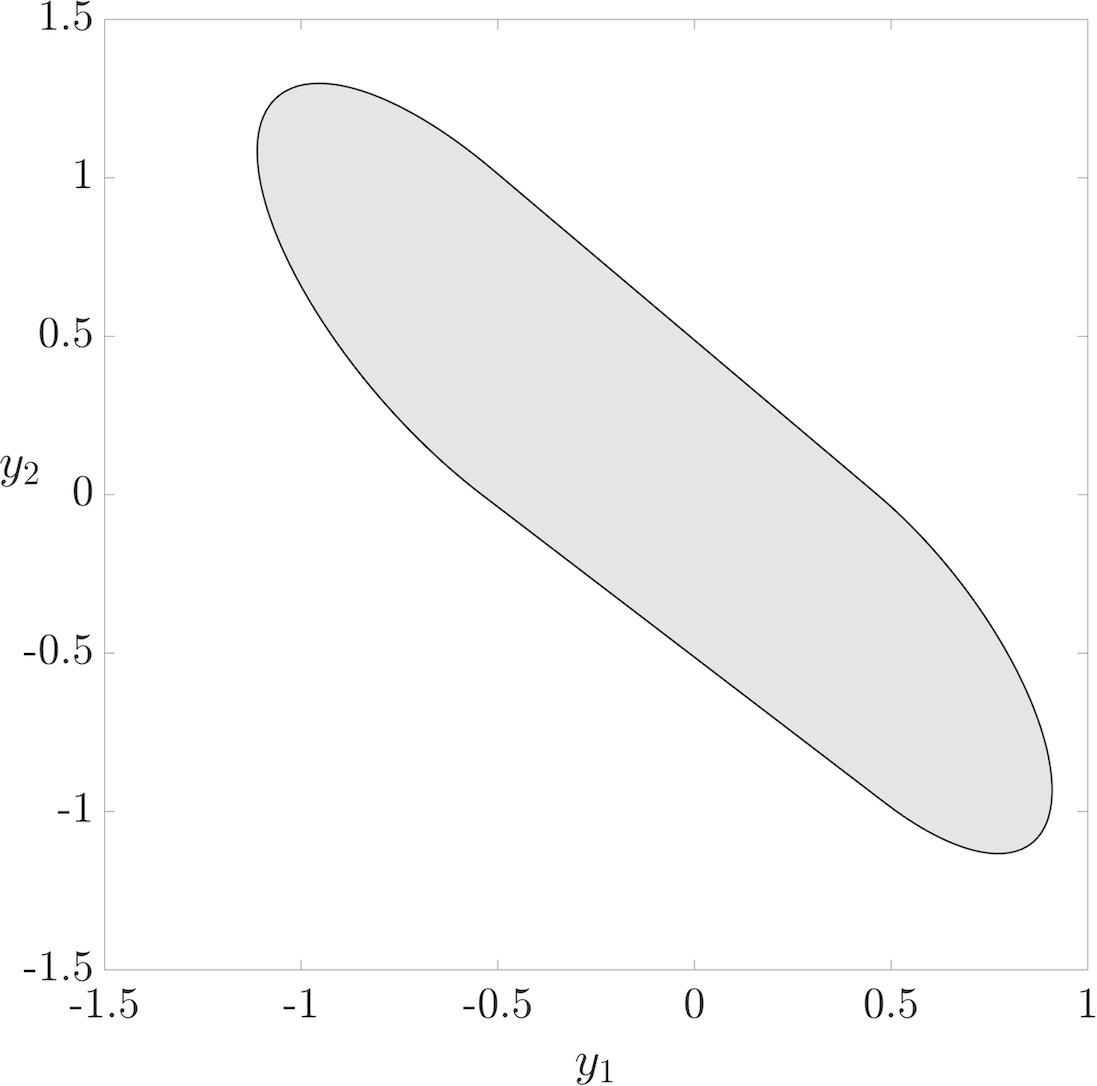}
  \caption{{\small{$\mathcal{R}^{\circ}$ with $\bm{x}_{0}=(0.05,0.05)^{\top}$.}}}
  \label{fig:sfig2}
\end{subfigure}
%\par\medskip % force a bit of vertical whitespace
%\begin{subfigure}{.24\textwidth}
%  \centering
%  \includegraphics[width=.99\linewidth]{2DIntegratorReachSetBigIC.png}
%  \caption{{\small{$\mathcal{R}$ with $\bm{x}_{0}=(0.5,0.5)^{\top}$.}}}
%  \label{fig:sfig3}
%\end{subfigure}
%\begin{subfigure}{.24\textwidth}
%  \centering
%  \includegraphics[width=.99\linewidth]{PolarDualBigIC.png}
%  \caption{{\small{$\mathcal{R}^{\circ}$ with $\bm{x}_{0}=(0.5,0.5)^{\top}$.}}}
%  \label{fig:sfig4}
%\end{subfigure}
\caption{{\small{The double integrator reach set $\mathcal{R}\left(\{\bm{x}_{0}\},t\right)$ and its polar dual $\left(\mathcal{R}\left(\{\bm{x}_{0}\},t\right)\right)^{\circ}$ at $t=2$, $\mathcal{U}\equiv[\alpha,\beta]=[-1,1]$. The curves $p^{\text{upper}},p^{\text{lower}}$ defining the reach set boundary (see Corollary \ref{Corollary:TwoBoundingSurfaces} and the discussion thereafter) are shown too.}}}
\vspace*{-0.12in}
\label{fig:DoubleIntegratorReachSetAndItsDual}
\end{figure}

We also know from Sec. \ref{subsec:zonoid} that $\mathcal{R}^{\Box}\left(\{\bm{x}_{0}\},t\right)$ is a zonoid. However, the polar of a zonoid is not a zonoid in general \cite{schneider1975zonoids,lonke1997zonoids}, and we should not expect $\left(\mathcal{R}^{\Box}\left(\{\bm{x}_{0}\},t\right)\right)^{\circ}$ to be one. Fig. \ref{fig:DoubleIntegratorReachSetAndItsDual} shows $\mathcal{R}\left(\{\bm{x}_{0}\},t\right)$ and $\left(\mathcal{R}\left(\{\bm{x}_{0}\},t\right)\right)^{\circ}$ for the double integrator ($d=2$, $m=1$).

\subsection{Summary of Taxonomy}\label{subsec:classifiationsummary}
So far we explained that the compact convex set $\mathcal{R}^{\Box}\left(\{\bm{x}_{0}\},t\right)$ is semialgebraic, and a translated zonoid. Two well-known subclasses of convex semialgebraic sets are the \emph{spectrahedra} and the \emph{spectrahedral shadows}. The spectrahedra, a.k.a. \emph{linear matrix inequality (LMI) representable sets} are affine slices of the symmetric positive semidefinite cone. The spectrahedral shadows, a.k.a. \emph{lifted LMI or semidefinite representable sets} are the projections of spectrahedra. The spectrahedral shadows subsume the class of spectrahedra; e.g., the set $\{(x_1,x_2)\in\mathbb{R}^{2}\mid x_{1}^{4}+x_{2}^{4}\leq 1\}$ is a spectrahedral shadow but not a spectrahedron. The polar duals of spectrahedra are spectrahedral shadows \cite[Ch. 5, Sec. 5.5]{blekherman2012semidefinite}.

\begin{figure}[t]
\begin{subfigure}{0.235\textwidth}
  \centering
  \includegraphics[width=0.94\linewidth]{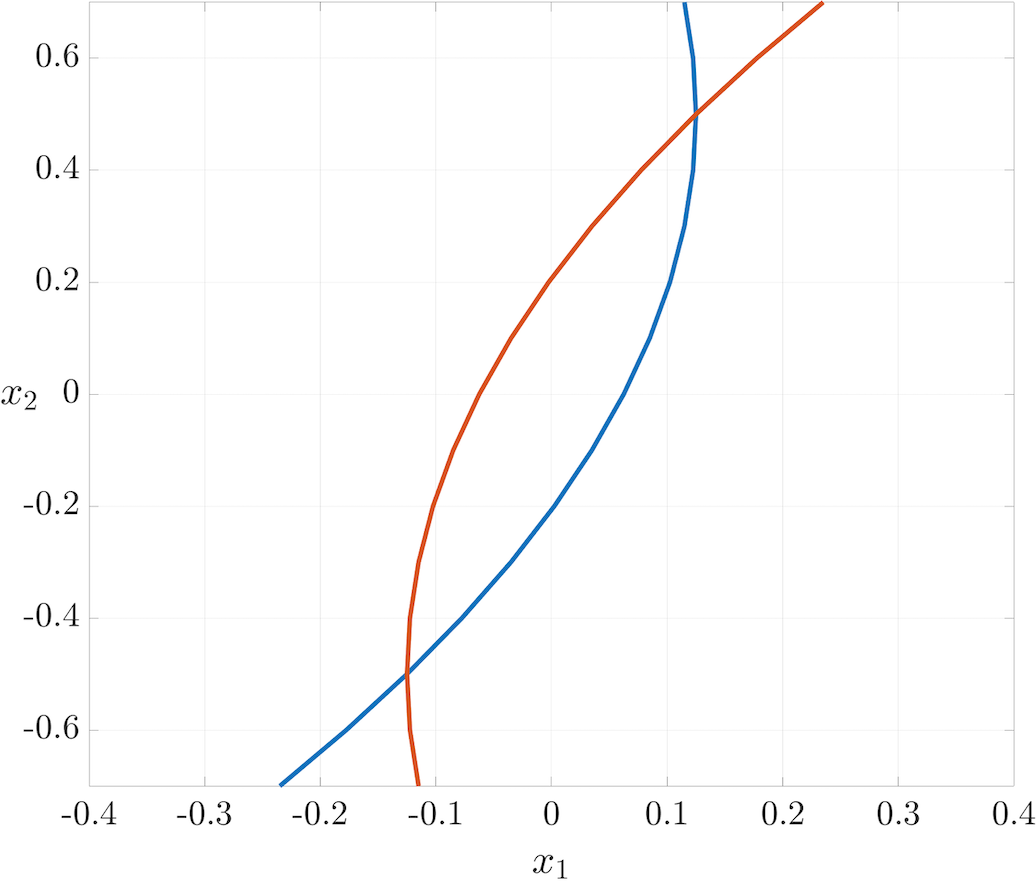}
  \caption{{\small{Real algebraic curves $p^{\text{upper}},\allowbreak p^{\text{lower}}$ for  the double integrator.}}}
  \label{fig:sfig1}
\end{subfigure}\hspace*{0.15in}
\begin{subfigure}{.235\textwidth}
  \centering
  \includegraphics[width=\linewidth]{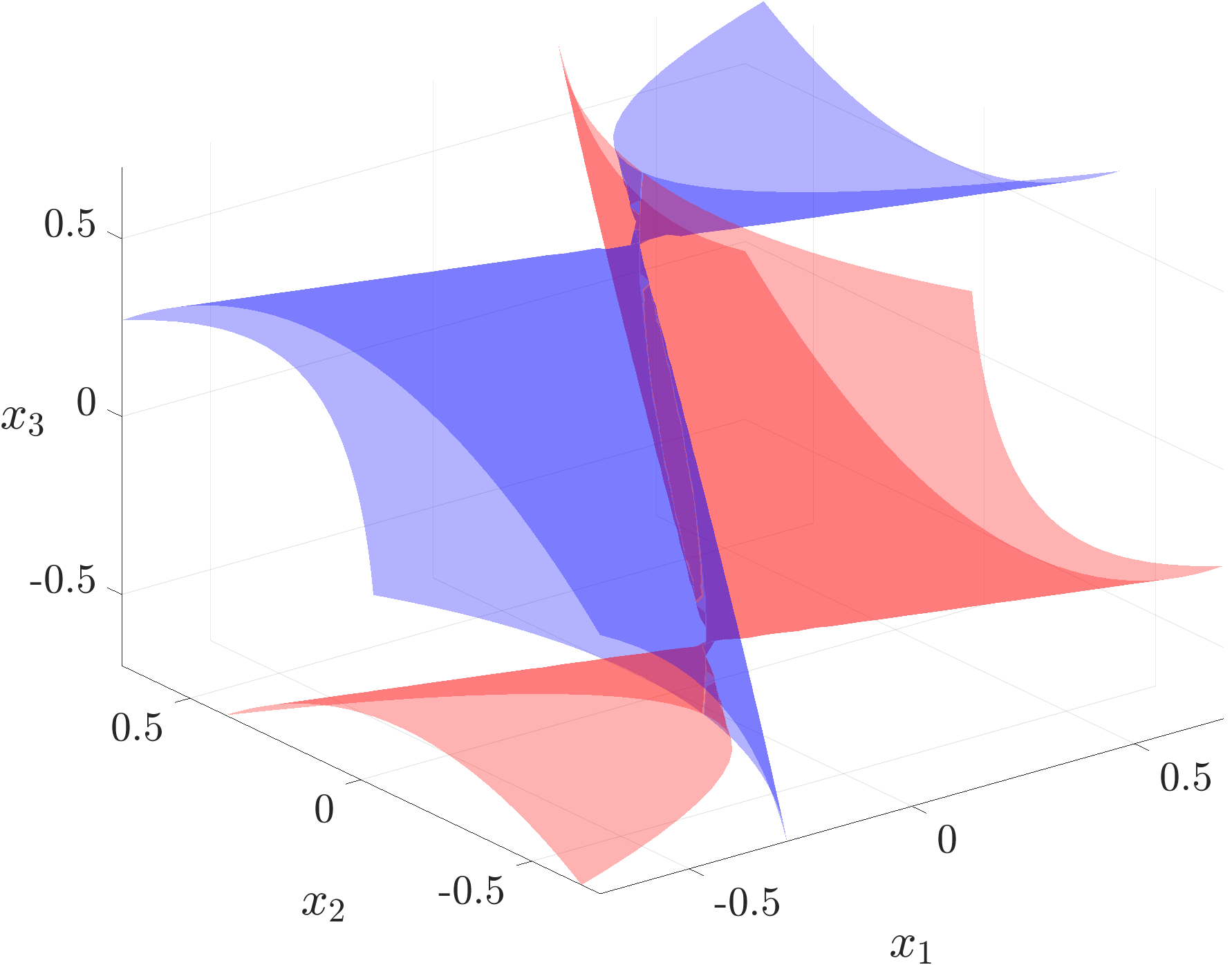}
  \caption{{\small{Real algebraic surfaces $p^{\text{upper}},\allowbreak p^{\text{lower}}$ for the triple integrator.}}}
  \label{fig:sfig2}
\end{subfigure}
\caption{{\small{The bounding polynomials for the double and triple integrator reach sets at $t=0.5$ with $\bm{x}_{0}=\bm{0}$ and $\mu=1$.}}}
\label{fig:Boundaries}
\vspace*{-0.2in}
\end{figure}

We note that $\mathcal{R}^{\Box}$ is not a spectrahedron. To see this, we resort to the contrapositive of \cite[Thm. 3.1]{helton2007linear}. Specifically, the number of intersections made by a generic line passing through an interior point of the $d$-dimensional reach set $\mathcal{R}^{\Box}$ with its real algebraic boundary is not equal to the degree of the bounding algebraic hypersurfaces, the latter we know from Sec. \ref{subsec:Implicitization} to be $(\lfloor \frac{d-1}{2}\rfloor + 1)(d-\lfloor \frac{d-1}{2}\rfloor)$. In other words, the $\mathcal{R}^{\Box}$ is not rigidly convex, see \cite[Sec. 3.1 and 3.2]{helton2007linear}. Fig. \ref{fig:Boundaries} helps visualize this for $m=1$. From Fig. \ref{fig:sfig1}, we observe that a generic line for $d=2$ has $4$ intersections with the bounding real algebraic curves whereas from (\ref{p1Implicit}), we know that $p^{\text{upper}},p^{\text{lower}}$ are degree $2$ polynomials. Likewise, Fig. \ref{fig:sfig2} reveals that a generic line for $d=3$ has $6$ intersections with the bounding real algebraic surfaces whereas from (\ref{p2Implicit}), we know that the polynomials $p^{\text{upper}},p^{\text{lower}}$ in this case, are of degree $4$.

\begin{figure}[t]
        \centering
        \includegraphics[width=0.9\linewidth]{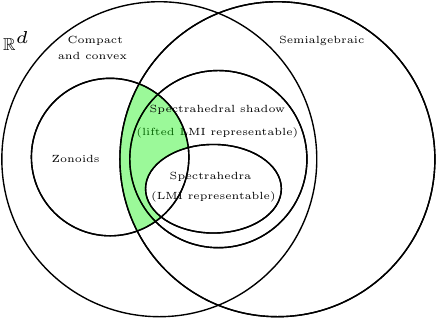}
        \caption{{\small{The summary of taxonomy for the integrator reach set $\mathcal{R}^{\Box}$.}}}
\vspace*{-0.2in}
\label{fig:taxonomysummary}
\end{figure}

Could the reach set $\mathcal{R}^{\Box}$ be spectrahedral shadow? Some calculations show that \emph{sufficient} conditions as in \cite{helton2010semidefinite} do not seem to hold. However, this remains far from conclusive. We summarize our taxonomy results in Fig. \ref{fig:taxonomysummary}; the highlighted region shows where the integrator reach set belongs. To answer whether this highlighted region can be further narrowed down, seems significantly more challenging.

%%%%%%%%%%%%%%%%%%%%%%%%%%%%%%%%%%%%%%%%%%%%%%%%%%%%%%%%%%%%%%%%%%%%%%%%%%%%%%%%%%%%%%%%%%%%%%%%%%%%%%%%%%%%%%%%%%%%%%%%%%%%%%%%%%%%%%%%%%%%%%%%%%%%%%%%%%%

\section{Size}\label{sec:Size}
%Notice however, that the (space-time) \emph{forward reachable tube} 
%\begin{align}
%\overline{\mathcal{R}}\left(\mathcal{X}_{0},t\right) := \bigcup_{0\leq \tau \leq t} \mathcal{R}\left(\mathcal{X}_{0},\tau\right),
%\label{DefReachableTube}	
%\end{align}
We next quantify the ``size" of the reach set $\mathcal{R}^{\Box}\left(\{\bm{x}_{0}\},t\right)$ by computing two functionals: its $d$-dimensional volume (Sec. \ref{subsec:Volume}), and its diameter or maximum width (Sec. \ref{subsec:Diameter}). In Sec. \ref{subsec:Scaling}, we discuss how these functionals scale with the state dimension $d$.

\subsection{Volume}\label{subsec:Volume}
The following result gives the volume formula for the integrator reach set $\mathcal{R}^{\Box}$.
\begin{theorem}\label{ThmVolIntegratorReachSet} 
Fix $\bm{x}_{0}\in\mathbb{R}^{d}$, let $\mathcal{X}_{0} \equiv \{\bm{x}_{0}\}$ and $\mathcal{U}$ given by \eqref{BoxInputSet}. Consider the integrator dynamics (\ref{IntegratorDyn})-(\ref{defAB}) with $d$ states, $m$ inputs, and relative degree vector $\bm{r}=(r_1,r_2,...,r_m)^{\top}$. Define $\mu_{1},\hdots,\mu_{m}$ as in (\ref{defalphabeta})-(\ref{defmujnuj}). Then the $d$-dimensional volume of the integrator reach set (\ref{DefReachSet}) at time $t>0$ is
\begin{align}
\vol\!\left(\mathcal{R}^{\Box}\left(\{\bm{x}_{0}\},t\right)\right) \!=\! 2^{d}\displaystyle\prod_{j=1}^{m} \!\bigg\{\mu_{j}^{r_{j}} t^{r_{j}(r_j + 1)/2} \!\prod_{k=1}^{r_{j}-1}\!\dfrac{k!}{(2k+1)!}\bigg\}.
\label{VolumeFormula}	
\end{align}
\end{theorem}

\begin{figure}[t]
        \centering
        \includegraphics[width=0.9\linewidth]{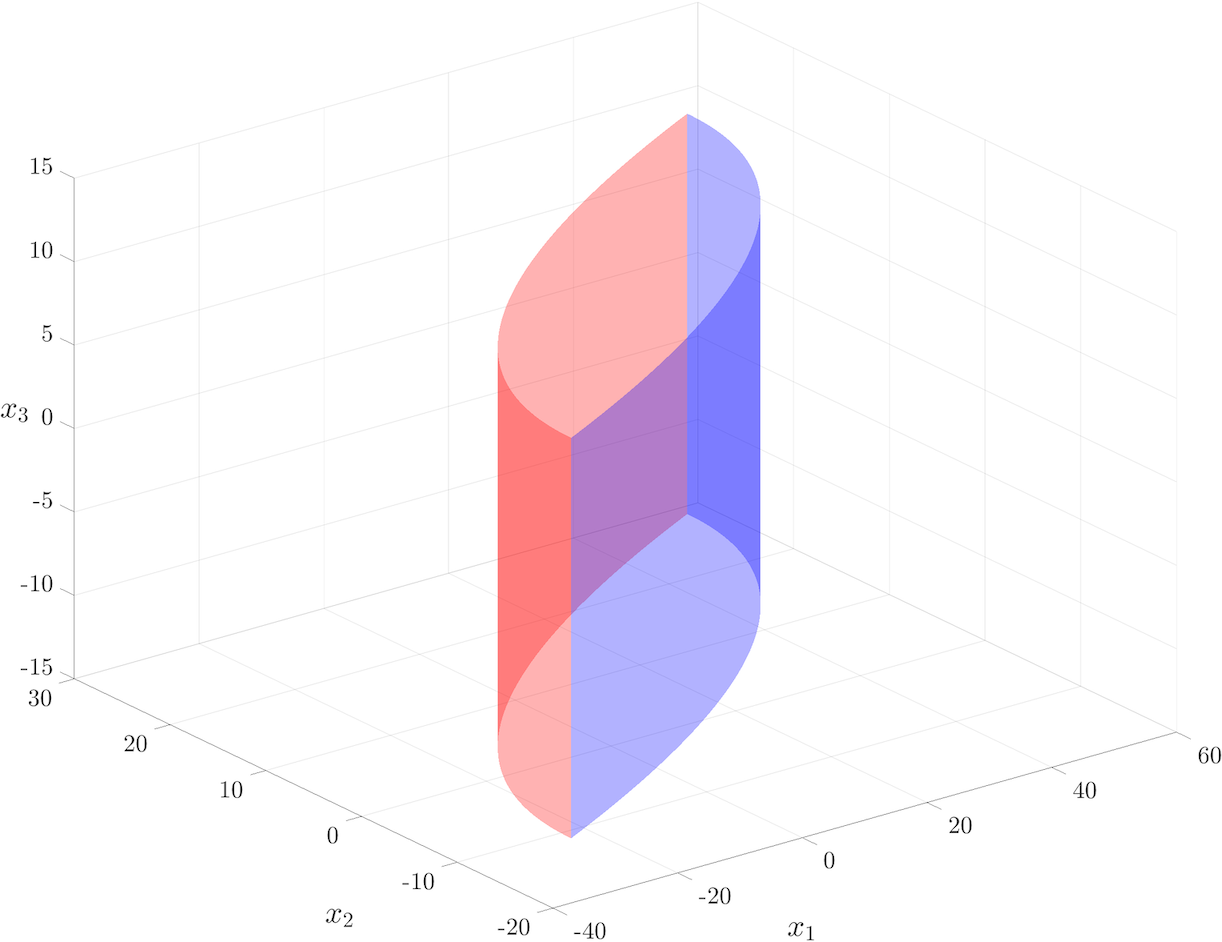}
        \caption{{\small{The integrator reach set $\mathcal{R}^{\Box}(\{\bm{x}_{0}\},t=4)$ with $m=2$, $\bm{r}=(2,1)^{\top}$, $\bm{x}_{0} = (1,1,0)^{\top}$, $[\alpha_{1},\beta_{1}]=[-5,5]$, $[\alpha_{2},\beta_{2}]=[-3,3]$.}}}
\vspace*{-0.2in}
\label{FigTwoInputReachSet3D}
\end{figure}

For a simple illustration of Theorem \ref{ThmVolIntegratorReachSet}, consider $d=3$, $m=2$ with $\bm{r}=(2,1)^{\top}$. The corresponding reach set $\mathcal{R}^{\Box}(\{\bm{x}_{0}\},t)$ at $t=4$ is shown in Fig. \ref{FigTwoInputReachSet3D} for $\bm{x}_{0} = (1,1,0)^{\top}$, $\mathcal{U}=[-5\times 5]\times[-3,3]$. Here $\mu_{1}=5$ and $\mu_{2}=3$.

This reach set, being a direct product of the double integrator reach set $\mathcal{R}_{1}$ (cf. Fig. \ref{fig:DoubleIntegratorReachSetAndItsDual}) and the single integrator reach set $\mathcal{R}_{2} = \{\bm{x}_{0}(3)\} \dotplus [-\mu_{2}t,\mu_{2}t]$, is a cylinder\footnote{Here, the notation $\bm{x}_{0}(3)$ stands for the third component of vector $\bm{x}_{0}$.}. In \cite{haddad2020convex}, we explicitly derived that $\vol\left(\mathcal{R}_{1}\right)= \frac{2}{3}\mu_{1}^2 t^3$, and therefore, the volume of this cylindrical set must be equal to {{``height of the cylinder $\times$ cross sectional area", i.e.,}}
\begin{align*}
2\mu_{2}t \times \frac{2}{3}\mu_{1}^2 t^3 = \frac{4}{3}\mu_{1}^{2}\mu_{2}t^{4}.	
\end{align*}
Indeed, a direct application of the formula (\ref{VolumeFormula}) recovers the above expression.

\begin{remark}\label{remark:VolNonsingletonX0}
If the initial set $\mathcal{X}_{0}$ is not singleton, then computing the volume of the $\mathcal{R}^{\Box}$ requires us to compute the volume of a Minkowski sum. Notice that 
\begin{align*}
\vol\left(\exp(t\bm{A})\mathcal{X}_{0}\right) &= |\det\left(\exp(t\bm{A})\right)|\vol\left(\mathcal{X}_{0}\right) \\
&= \exp\left(\tr\left(t\bm{A}\right)\right)\vol\left(\mathcal{X}_{0}\right)\\
&= \exp\left(\!\sum_{j=1}^{m}\tr\left(t\bm{A}_{j}\right)\!\!\right)\vol\left(\mathcal{X}_{0}\right) \\
&= \vol\left(\mathcal{X}_{0}\right),
\end{align*}
since from (\ref{defBlocks}), $\tr(\bm{A}_{j})=0$ for all $j=1,\hdots,m$. Therefore, combining (\ref{SetValuedIntegral}), (\ref{VolumeFormula}) with the classical Brunn-Minkowski inequality, we obtain a lower bound for $\vol\left(\mathcal{R}^{\Box}\right)$ as
\begin{align*}
&\left(\vol\left(\mathcal{R}^{\Box}\left(\mathcal{X}_{0},t\right)\right)\right)^{1/d} \geq \left(\vol\left(\mathcal{X}_{0}\right)\right)^{1/d} \\
& \qquad\qquad + 2\left(\displaystyle\prod_{j=1}^{m} \bigg\{\mu_{j}^{r_{j}} t^{r_{j}(r_j + 1)/2} \prod_{k=1}^{r_{j}-1}\dfrac{k!}{(2k+1)!}\bigg\}\right)^{\!\!1/d}.	
\end{align*}
The above bound holds for any compact $\mathcal{X}_{0}\subset\mathbb{R}^{d}$, not necessarily convex.  
\end{remark}

%%%%%%%%%%%%%%%%%%%%%%%%%%%%%%%%%%%%%%%%%%%%%%%%%%%%%%%%%%%%%%%%%%%%%%%%%%%%%

\subsection{Diameter}\label{subsec:Diameter}
We now focus on another measure of ``size" for the integrator reach set $\mathcal{R}^{\Box}$, namely its diameter, or maximal width. 

By definition, the \emph{width} \cite[p. 42]{schneider2014convex} of $\mathcal{R}^{\Box}\left(\mathcal{X}_{0},t\right)$, is 
\begin{align}
w_{\mathcal{R}^{\Box}\left(\mathcal{X}_{0},t\right)}(\bm{\eta}) := h_{\mathcal{R}^{\Box}(\mathcal{X}_{0},t)}\left(\bm{\eta}\right) + h_{\mathcal{R}^{\Box}(\mathcal{X}_{0},t)}\left(-\bm{\eta}\right),
\label{DefWidth}	
\end{align}
where $\bm{\eta}\in\mathbb{S}^{d-1}$ (the unit sphere imbedded in $\mathbb{R}^{d}$), and the support function $h_{\mathcal{R}^{\Box}(\mathcal{X}_{0},t)}\left(\cdot\right)$ is given by (\ref{SptFnIntegratorFinal}). In other words, (\ref{DefWidth}) gives the width of $\mathcal{R}^{\Box}$ in the direction $\bm{\eta}$. 

For singleton $\mathcal{X}_{0} \equiv \{\bm{x}_{0}\}$, combining (\ref{SptFnIntegratorFinal}) and (\ref{DefWidth}), we have
\begin{align}
w_{\mathcal{R}^{\Box}\left(\{\bm{x}_{0}\},t\right)}(\bm{\eta}) &= \int_{0}^{t} \bigg\{\lvert\langle\bm{\eta},\bm{\xi}(s)\rangle\rvert + \lvert\langle-\bm{\eta},\bm{\xi}(s)\rangle\rvert\bigg\}\:\differential s \nonumber\\
	&= 2\int_{0}^{t} \lvert\langle\bm{\eta},\bm{\xi}(s)\rangle\rvert\:\differential s,\label{widthformula}
\end{align}
where the last equality follows from the fact that $\bm{\xi}(s)$ in (\ref{xiVector}) is component-wise nonnegative for all $0 \leq s \leq t$. 

The \emph{diameter} of the reach set $\mathcal{R}^{\Box}$ is its maximal width:
\begin{align}
{\rm{diam}}\left(\mathcal{R}^{\Box}\left(\mathcal{X}_{0},t\right)\right) := \underset{\bm{\eta}\in\mathbb{S}^{d-1}}{\max}\:w_{\mathcal{R}^{\Box}	\left(\mathcal{X}_{0},t\right)}(\bm{\eta}).
\label{DefDiam}	
\end{align}
Notice that (\ref{widthformula}) is a convex function of $\bm{\eta}$; see e.g., \cite[p. 79]{boyd2004convex}. Thus, computing (\ref{DefDiam}) amounts to maximizing a convex function over the unit sphere. We next derive a closed form expression for (\ref{DefDiam}).

\begin{theorem}\label{ThmDiamIntegratorReachSet}
Fix $\bm{x}_{0}\in\mathbb{R}^{d}$, let $\mathcal{X}_{0} \equiv \{\bm{x}_{0}\}$ and $\mathcal{U}$ given by \eqref{BoxInputSet}. Consider the integrator dynamics (\ref{IntegratorDyn})-(\ref{defAB}) with $d$ states, $m$ inputs, and relative degree vector $\bm{r}=(r_1,r_2,...,r_m)^{\top}$. Define $\mu_{1},\hdots,\mu_{m}$ as in (\ref{defalphabeta})-(\ref{defmujnuj}). The diameter of the integrator reach set (\ref{DefReachSet}) at time $t>0$ is
\begin{align}
{\rm{diam}}\!\left(\!\mathcal{R}^{\Box}\left(\{\bm{x}_{0}\},t\right)\!\right) \!=\! 2 \parallel\!\bm{\zeta}(t) \!\parallel_{2}\:= 2\!\left(\displaystyle \sum_{j=1}^{m} \mu_{j}^{2} \| \bm{\zeta}_{j} \|^{2}\!\right)^{\!\!\frac{1}{2}}
\label{expandedformDiamFormula}
\end{align}
wherein $\bm{\zeta}(t)$ is defined as in Sec. \ref{subsubsec:sptfn}, and the $i$th component of the subvector $\bm{\zeta}_{j}(t)\in\mathbb{R}^{r_j}$ is
\begin{align}
\!\displaystyle\int_{0}^{t}\!\!\dfrac{s^{(r_j-i)}}{(r_j-i)!}\:\differential s= \dfrac{t^{r_j-i+1}}{(r_j-i+1)!}, \quad i=1,2,\hdots,r_{j}.
\end{align}
\end{theorem}

To illustrate Theorem \ref{ThmDiamIntegratorReachSet}, consider the triple integrator with {\small {$d=3$}} and {\small {$m=1$}}. In this case, $\mathcal{U}=[\alpha,\beta]$, $\mu:=(\beta-\alpha)/2$, and we can parameterize the unit vector $\bm{\eta}\in\mathbb{S}^{2}$ as
\[\bm{\eta} \equiv \begin{pmatrix}
\sin\theta\cos\phi\\
\sin\theta\sin\phi\\
\cos\theta
\end{pmatrix}\displaystyle, \quad \theta\in[0,\pi], \quad \phi\in[0, 2\pi).\] 
Thus (\ref{DefDiam}) reduces to
\[2\mu\:\underset{\substack{\theta\in[0,\pi]\\ \phi\in[0,2\pi)}}{\max}\:\displaystyle\int_{0}^{t}\lvert s^2\left(\sin\theta\cos\phi\right)/2 + s\sin\theta\sin\phi+\cos \theta \rvert\:\differential s.\]
Furthermore, $\bm{\zeta}(t) = (t^{3}/6,t^2/2, t)^{\top}$, and we obtain
\begin{align*}
\bm{\eta}^{\max} = \begin{pmatrix}
\!\sin\theta^{\max}\sin\phi^{\max}\!\\
\!\sin\theta^{\max}\cos\phi^{\max}\!\\
\!\cos\theta^{\max}\!	
\end{pmatrix}
= \dfrac{\pm1}{\sqrt{t^{4}+9t^{2}+36}}\begin{pmatrix}t^{2}\\
3t\\
6\end{pmatrix},	
\end{align*}
where $\pm$ means that either all components are plus or all minus.
Thus, the maximizing tuples $\left(\phi^{\max},\theta^{\max}\right) \in [0,\pi] \times [0, 2\pi)$ are given by
\begin{align}
&\left(\phi^{\max},\theta^{\max}\right) \nonumber\\
&= \begin{cases}
\left(\arctan\left(3/t\right), \arccos \left(6/\sqrt{t^{4}+9t^{2}+36}\right) \right),\\
\left(\pi+\arctan\left(3/t\right), \arccos\left(-6/\sqrt{t^{4}+9t^{2}+36}\right) \right).
\end{cases}
\label{TripleIntegratorMaximizers}
\end{align}
Hence, the diameter of the triple integrator reach set at time $t$ is equal to $\left(\mu t/3\right)\sqrt{t^{4} + 9t^{2} + 36}$.

\begin{figure}[t]
        \centering
        \includegraphics[width=\linewidth]{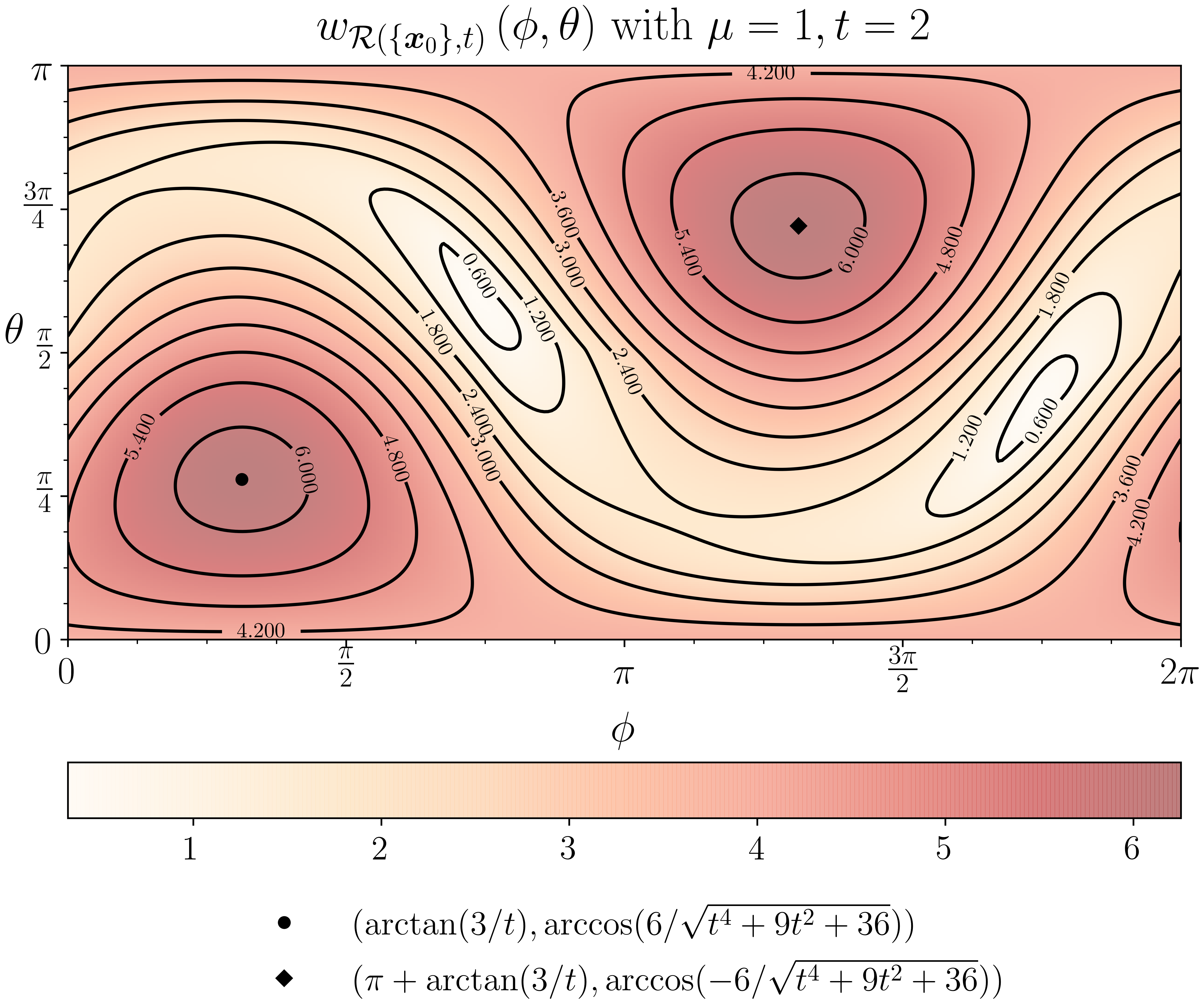}
        \caption{{\small{The width (\ref{widthformula}) for the single input triple integrator reach set $\mathcal{R}\left(\{\bm{x} _{0}\},t\right)$ is shown as a function of $(\phi,\theta)\in [0,\pi]\times[0,2\pi)$, which parameterize the unit sphere $\mathbb{S}^{2}$. Here $\mathcal{U}=[-1,1]$ and hence $\mu=1$. The darker (resp. lighter) hues correspond to the higher (resp. lower) widths. The filled black circle and the filled black square correspond to the maximizers $\left(\phi^{\max},\theta^{\max}\right)$ given by (\ref{TripleIntegratorMaximizers}).}}}
\vspace*{-0.2in}
\label{FigWidthReachSet}
\end{figure}

Fig. \ref{FigWidthReachSet} shows how the width of the integrator reach set for $d=3$, $m=1$ varies over $(\phi,\theta)\in [0,\pi]\times[0,2\pi)$, which parameterize the unit sphere $\mathbb{S}^{2}$.  The location of the  maximizers are given by (\ref{TripleIntegratorMaximizers}), and are depicted in Fig. \ref{FigWidthReachSet} via filled black circle and filled black square.

For a visualization of the width and diameter for the double integrator, see \cite[Fig. 2]{haddad2020convex}.

%%%%%%%%%%%%%%%%%%%%%%%%%%%%%%%%%%%%%%%%%%%%%%%%%%%%%%%%%%%%%%%%%%%%%%%%%%%%%

\subsection{Scaling Laws}\label{subsec:Scaling}
We now turn to investigate how the volume and the diameter of the integrator reach set scale with time and the state dimension. While scaling laws reveal limits of performance of engineered systems (see e.g., \cite{xie2004network,xue2006scaling}), they have not been formally investigated in the context of reach sets even though empirical studies are common \cite{althoff2011zonotope}, \cite[Table 1]{fan2016locally}, \cite[Fig. 4]{meng2022learning}.

For clarity, we focus on the single input case and hence do not notationally distinguish between $\mathcal{R}$ and $\mathcal{R}^{\Box}$.
%%%%%%%%%%%%%%%%%%%%%%%%%%%
\subsubsection{Scaling of the volume}
\begin{figure}[t]
        \centering
        \includegraphics[width=0.95\linewidth]{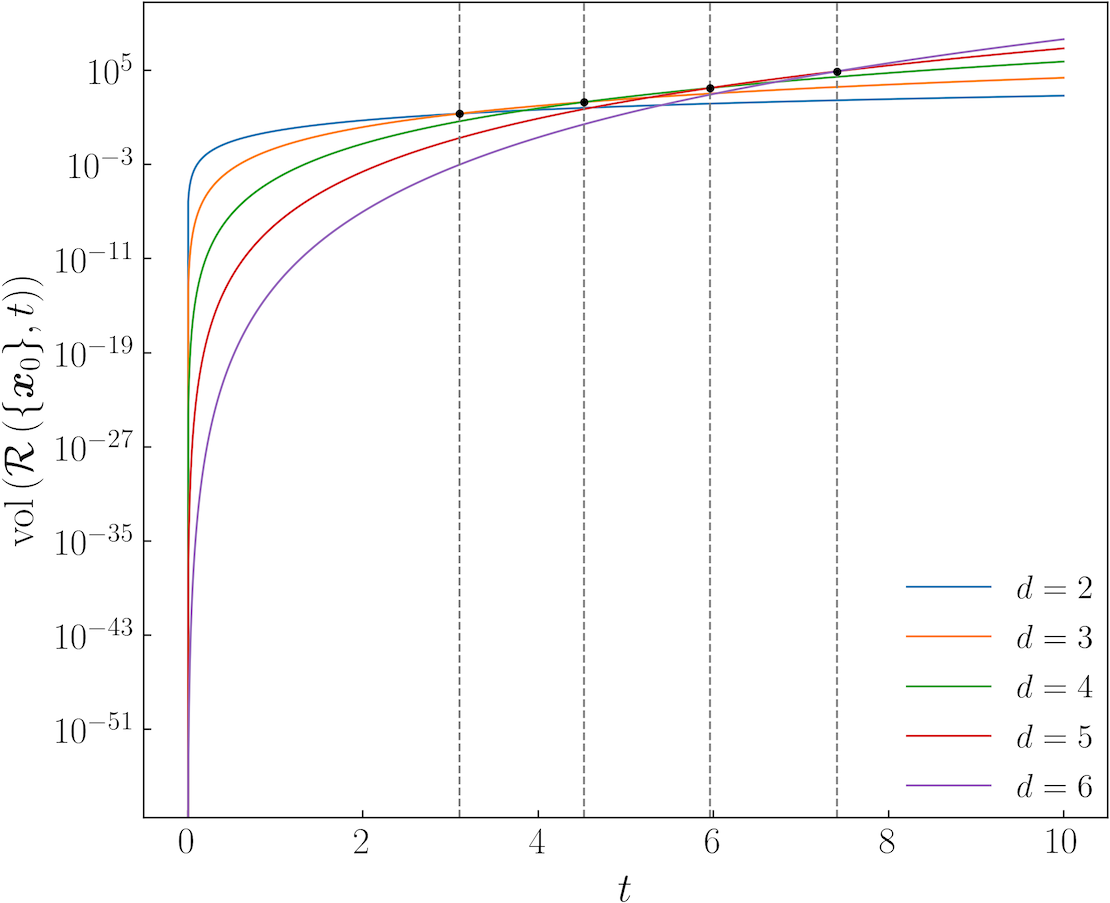}
        \caption{{\small{For single input ($m=1$), the volume of the integrator reach set $\mathcal{R}\left(\{\bm{x} _{0}\},t\right)\displaystyle$ computed from (\ref{VolumeFormula}) is plotted against time $t\displaystyle$ for state dimensions $d=2,3,\hdots,6$ with $\mathcal{U}=[\alpha,\beta]=[-1,1]$, $\mu:=(\beta-\alpha)/2=1$. The dashed vertical lines show the critical times given by (\ref{tcritical_d_delta}).}}}
\vspace*{-0.2in}
\label{FigVolumeTime}
\end{figure}

Fig. \ref{FigVolumeTime} plots the volume (\ref{VolumeFormula}) for the single input ($m=1$) case against time $t$ for varying state space dimension $d$. In this case, $\mathcal{U}=[\alpha,\beta]$, and therefore $\mu:=(\beta-\alpha)/2$. As expected, the volume of the reach set increases with time for any fixed $d$.

Let us now focus on the scaling of the volume with respect to the state dimension $d$. For $m=1$, using the known asymptotic \cite{oeisA107254} for $\prod_{k=1}^{d-1}(2k+1)!/k!$, we find the $d\rightarrow\infty$ asymptotic for the volume: 
\[\vol\left(\mathcal{R}_{d}\left(\{\bm{x}_{0}\},t\right)\right) \sim (2\mu)^{d}t^{d(d+1)/2}\dfrac{\exp\left(\frac{3}{2}d^2 + \frac{1}{12}\right)}{c\times 2^{\left(2d^{2}-\frac{1}{12}\right)} \: d^{\left(d^2 + \frac{1}{12}\right)}},\]
where $c \approx 1.2824\hdots$ is the Glaisher-Kinkelin constant \cite[Sec. 2.15]{finch2003mathematical}.

Fig. \ref{FigVolumeTime} shows that when $t$ is small, the volume of the larger dimensional reach set stays lower than its smaller dimensional counterpart. In particular, given two state space dimensions $d,d^{\prime}$ with $d>d^{\prime}$, and all other parameters kept fixed, there exists a critical time $t_{\text{cr}}$ when the volume of the $d$ dimensional reach set overtakes that of the $d^{\prime}$ dimensional reach set.

For any $d>d^{\prime}$, the critical time $t_{\text{cr}}$ satisfies
\begin{align*}
\underbrace{\vol\left(\mathcal{R}_{d}\left(\{\bm{x}_{0}\},t_{\text{cr}}\right)\right)}_{d\;\text{dimensional volume}}=\underbrace{\vol\left(\mathcal{R}_{d^{\prime}}\left(\{\bm{x}_{0}\},t_{\text{cr}}\right)\right)}_{d^{\prime}\;\text{dimensional volume}},
\end{align*}
which together with (\ref{VolumeFormula}) yields
\begin{align}
&t_{\text{cr}}=\left(2\mu\right)^{-\frac{2}{d+d^{\prime}+1}}\left(  \prod_{k=d^{\prime}}^{d-1} \dfrac{\left( 2k+1\right)!}{k!}\right)^{\frac{2}{(d-d^{\prime})(d+d^{\prime}+1)}}.
\end{align}

In particular, for $d^{\prime}=d-1$, we get
\begin{align}
t_{\text{cr}}=\left(\frac{1}{2\mu}\frac{(2d-1)!}{(d-1)!}\right)^{1/d}, \quad d=2,3,\hdots.
\label{tcritical_d_delta}
\end{align}
For instance, when $\mu=1$, $d=3$, $d^{\prime}=2$, we have $t_{\text{cr}} = (30)^{1/3} \approx 3.1072$. When $\mu=1$, $d=4$, $d^{\prime}=3$, we have $t_{\text{cr}} = (420)^{1/4} \approx 4.5270$. The dashed vertical lines in Fig. \ref{FigVolumeTime} show the critical times given by (\ref{tcritical_d_delta}).

Applying Stirling's approximation $n! \sim \sqrt{2\pi n} (n/e)^{n}$, we obtain the $d\rightarrow\infty$ asymptotic for (\ref{tcritical_d_delta}):
\begin{align*}
t_{\text{cr}} \sim \frac{4}{e}\:d\:\mu^{-\frac{1}{d}}\:2^{-\frac{3}{2d}},
\end{align*}
where $\sim$ denotes asymptotic equivalence \cite[Ch. 1.4]{de1981asymptotic}, and $e$ is the Euler number.

\subsubsection{Scaling of the diameter}
\begin{figure}[t]
        \centering
        \includegraphics[width=0.95\linewidth]{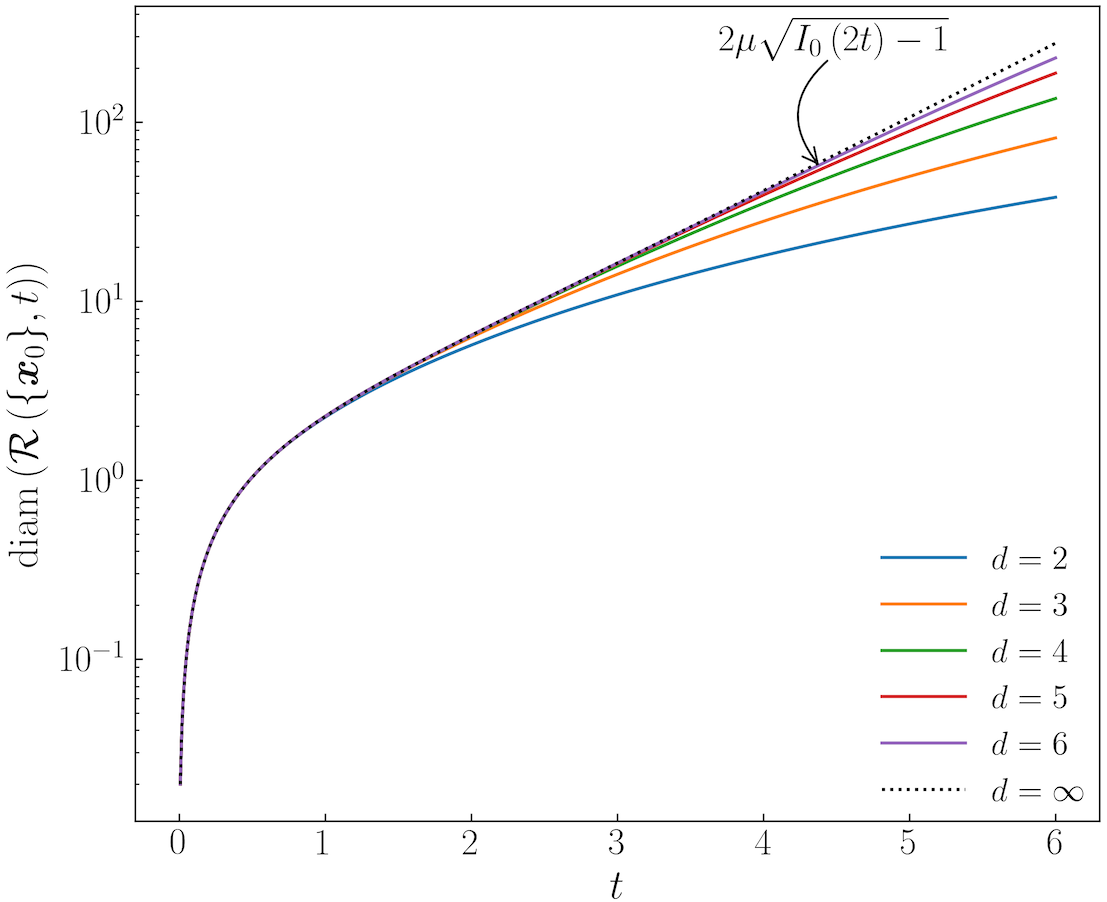}
        \caption{{\small{For single input ($m=1$), the diameter of the integrator reach set $\mathcal{R}\left(\{\bm{x} _{0}\},t\right)\displaystyle$ computed from (\ref{expandedformDiamFormula}) is plotted against time $t\displaystyle$ for state dimensions $d=2,3,\hdots,6$ with $\mathcal{U}=[\alpha,\beta]=[-1,1]$, $\mu:=(\beta-\alpha)/2=1$. As $d\to\infty\displaystyle$, the diameter converges to $2\mu\sqrt{I_0(2t)-1}\displaystyle$, shown by the dotted line.}}}
\vspace*{-0.2in}
\label{FigWidthTime}
\end{figure}

Fig. \ref{FigWidthTime} plots the diameter of (\ref{expandedformDiamFormula}) for the single input ($m=1$) case against time $t$ for varying state space dimension $d$. As earlier, $\mathcal{U}=[\alpha,\beta]$, $\mu:=(\beta-\alpha)/2$. As expected, the diameter of the reach set increases with time for any fixed $d$. 

As $d\rightarrow\infty$, the diameter approaches a limiting curve shown by the dotted line in Fig. \ref{FigWidthTime}. To derive this limiting curve, notice that for $m=1$, the formula (\ref{expandedformDiamFormula}) gives
\begin{align}
\displaystyle\lim_{d\rightarrow\infty} {\rm{diam}}\left(\mathcal{R}\left(\{\bm{x}_{0}\},t\right)\right) = \lim_{d \to \infty} 2\mu \sqrt{\sum_{j=1}^{d}\left(\dfrac{t^j}{j!}\right)^2}.
\label{DiamLimit}
\end{align}
We write the partial sum
\begin{align}
\sum_{j=1}^{d}\left(\dfrac{t^j}{j!}\right)^{2} = \underbrace{\sum_{j=1}^{\infty}\left(\dfrac{t^j}{j!}\right)^{2}}_{=: S_{1}} - \underbrace{\sum_{j=d+1}^{\infty}\left(\dfrac{t^j}{j!}\right)^{2}}_{=: S_{2}},
\label{RewritePartialSum}
\end{align}
and by ratio test, note that both the sums $S_{1},S_{2}$ converge. In particular, $S_{1}$ converges to $I_{0}(2t) - 1$, where $I_{0}(\cdot)$ is the zeroth order modified Bessel function of the first kind. This follows from the very definition of the $\nu$th order modified Bessel function of the first kind, given by
\[I_{\nu}(z) := \left(z/2\right)^{\nu} \displaystyle\sum_{j=0}^{\infty} \frac{\left(z^{2}/4\right)^{j}}{j!\:\Gamma\left(\nu+j+1\right)}, \quad \nu\in\mathbb{R},\]
where $\Gamma(\cdot)$ denotes the Gamma function.

On the other hand, using the definition of the generalized hypergeometric function\footnote{Here, $(\cdot)_{n}$ denotes the Pochhammer symbol \cite[p. 256]{abramowitz1970handbook} or rising factorial.} 
\[ _{1}F_{2}\left(a_{1};b_{1},b_{2};z\right) := \sum_{n=0}^{\infty}\dfrac{(a_1)_{n}}{(b_1)_{n}(b_{2})_{n}}\dfrac{z^{n}}{n!},\]
we find that   
\[S_{2} = \dfrac{t^{2(d+1)}\:_{1}F_{2}\left(1;d+2,d+2;t^{2}\right)}{\left((d+1)!\right)^{2}}.\]
Therefore, (\ref{RewritePartialSum}) evaluates to
\begin{align}
\!\!\!S_{1}\!-\!S_{2} = I_{0}(2t)-1 - \dfrac{t^{2(d+1)}\:_{1}F_{2}\left(1;d+2,d+2;t^{2}\right)}{\left((d+1)!\right)^{2}}\!.
\label{S1minusS2}	
\end{align}

Combining (\ref{DiamLimit}), (\ref{RewritePartialSum}), (\ref{S1minusS2}), and using the continuity of the square root function on $[0,\infty)$, we deduce that
\begin{align}
\!\!\!\!\!\displaystyle\lim_{d\rightarrow\infty} {\rm{diam}}\left(\mathcal{R}\left(\{\bm{x}_{0}\},t\right)\right) &= 2\mu\sqrt{\lim_{d \to \infty}\left(S_1 - S_2\right)} \nonumber\\
&= 2\mu\sqrt{I_{0}(2t) - 1}.
\label{LimitingDiam}	
\end{align}
That $\lim_{d\rightarrow\infty}S_{2}$ exists and equals to zero, follows from (\ref{RewritePartialSum}) and the continuity of the square:
\[\lim_{d\rightarrow\infty}S_{2} = \lim_{j\rightarrow\infty}\left(\frac{t^{j}}{j!}\right)^{\!\!2} = \left(\lim_{j\rightarrow\infty}\frac{t^{j}}{j!}\right)^{\!\!2}=0.\]
To see the last equality, let $a_{j}:=t^{j}/j!$. By the ratio test, $\underset{j\rightarrow\infty}{\lim\sup}|a_{j+1}/a_{j}| = \underset{j\rightarrow\infty}{\lim}t/j = 0 < 1$, hence $\{a_{j}\}$ is a Cauchy sequence and $\underset{j\rightarrow\infty}{\lim}a_{j}=0$.

The dotted line in Fig. \ref{FigWidthTime} is the curve (\ref{LimitingDiam}). 

%%%%%%%%%%%%%%%%%%%%%%%%%%%%%%%%%%%%%%%%%%%%%%%%%%%%%%%%%%%%%%%%%%%%%%%%%%%%%
%%%%%%%%%%%%%%%%%%%%%%%%%%%%%%%%%%%%%%%%%%%%%%%%%%%%%%%%%%%%%%%%%%%%%%%%%%%%%%%%

\section{Benchmarking Over-approximations of Integrator Reach Sets}\label{sec:Applications}
In practice, a standard approach for safety and performance verification is to compute ``tight" over-approximation of the reach sets of the underlying controlled dynamical system. Several numerical toolboxes such as \cite{kurzhanskiy2006ellipsoidal,althoff2015introduction} are available which over-approximate the reach sets using simple geometric shapes such as zonotopes and ellipsoids. Depending on the interpretation of the qualifier ``tight", different optimization problems ensue, e.g., minimum volume outer-approximation \cite{schweppe1973uncertain,chernousko1980optimal,chernousko1981guaranteed,maksarov1996state,durieu2001multi,alamo2005guaranteed,halder2018parameterized,halder2020smallest}.

\begin{figure}[t]
        \centering
        \includegraphics[width=0.97\linewidth]{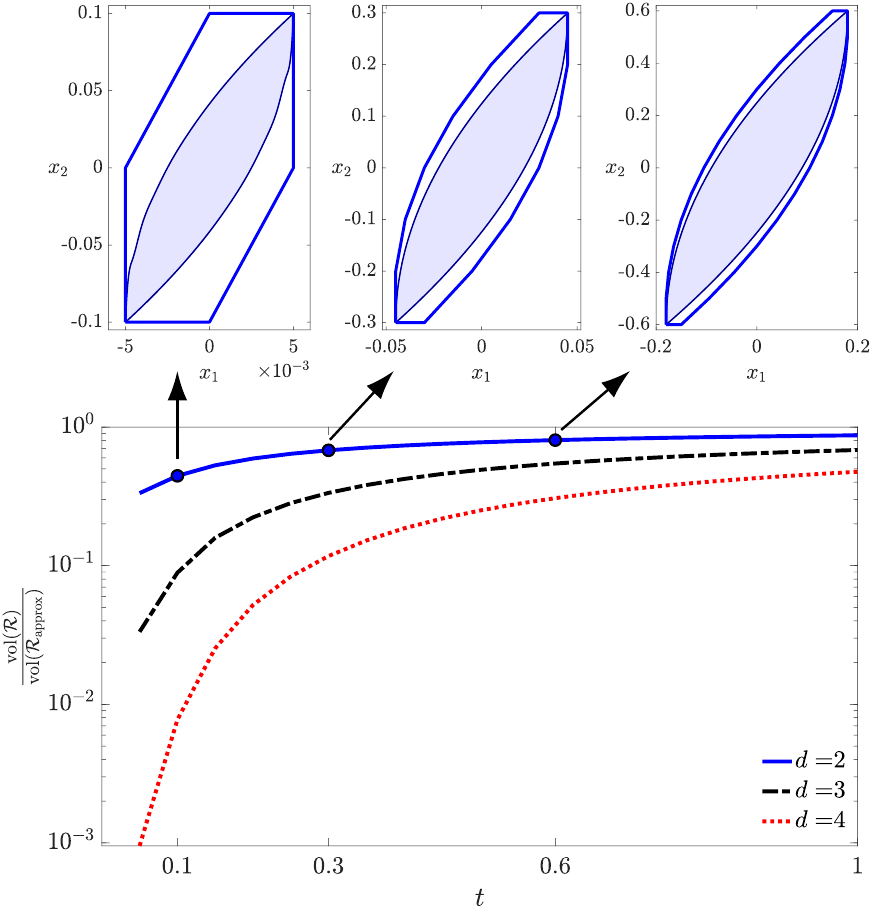}
        \caption{{\small{\emph{(Top)} Zonotopic over-approximations of the double integrator reach sets; \emph{(bottom)} the ratio of the volume of the single input integrator reach set $\mathcal{R}(t)$ and that of its zonotopic over-approximation $\mathcal{R}_{\text{approx}}(t)$ for $d=2,3,4$, plotted against time $t\in[0,1]$. The results are computed using the CORA toolbox with $\mu=1$, $\mathcal{X}_{0}=\{\bm{0}\}$.}}}
\vspace*{-0.2in}
\label{ZonotopeOuterApprox}
\end{figure}

One potential application of our results in Sec. \ref{sec:Size} is to help quantify the conservatism in different over-approximation algorithms by taking the integrator reach set as a benchmark case. For instance, Fig. \ref{ZonotopeOuterApprox} shows the conservatism in zonotopic over-approximations $\mathcal{R}_{\text{approx}}(t)$ of the single input integrator reach sets $\mathcal{R}(\{\bm{0}\},t)\subseteq\mathcal{R}_{\text{approx}}(\{\bm{0}\},t)$ for $d=2,3,4$ with $0\leq t \leq 1$ and $\mu=1$, computed using the CORA toolbox \cite{althoff2015introduction,CORAgithub}. To quantify the conservatism, we used the volume formula (\ref{VolumeFormula}) for computing the ratio of the volumes $\vol(\mathcal{R})/\vol(\mathcal{R}_{\text{approx}}) \in [0,1]$. The results shown in Fig. \ref{ZonotopeOuterApprox} were obtained by setting the zonotope order $50$ in the CORA toolbox, which means that the number of zonotopic segments used by CORA for over-approximation was $\leq 50 d$. As expected, increasing the zonotope order improves the accuracy at the expense of computational speed, but among the different dimensional volume ratio curves, trends similar to Fig. \ref{ZonotopeOuterApprox} remain. It is possible \cite[Thm. 1.1, 1.2]{bourgain1989approximation} to compute the optimal zonotope order as function of the desired approximation accuracy (i.e., desired Hausdorff distance from the zonoid).

For the numerical results shown in Fig. \ref{ZonotopeOuterApprox}, we found the diameters of the over-approximating zonotopes for $d=2,3,4$, to be the same as that of the true diameters given by (\ref{expandedformDiamFormula}) for all times.

\begin{figure*}[t]
        \centering
        \includegraphics[width=0.9\linewidth]{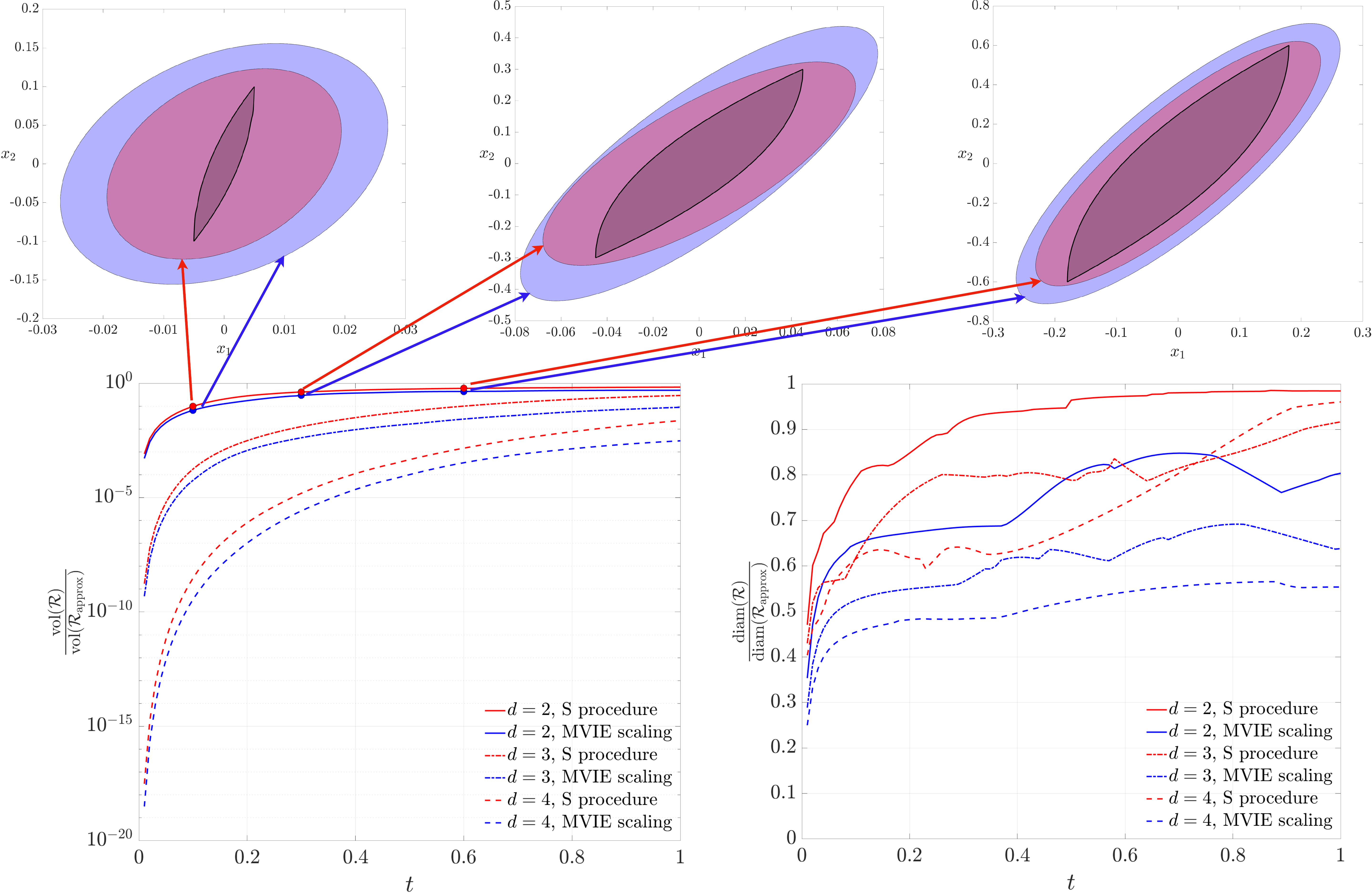}
        \caption{{\small{\emph{(Top)} Ellipsoidal over-approximations of the double integrator reach sets; \emph{(bottom)} the ratio of the volume (\emph{left}) and diameter (\emph{right}) of the single input integrator reach set $\mathcal{R}(t)$ and that of its ellipsoidal over-approximation $\mathcal{R}_{\text{approx}}(t)$ for $d=2,3,4$, plotted against time $t\in[0,1]$. Two different ellipsoidal over-approximations are shown: one \emph{(in red)} based on the S procedure, and the other \emph{(in blue)} obtained by scaling the maximum volume inner ellipsoid (MVIE) of the intersection of a parameterized family of ellipsoids. The results are computed for $\mu=1$, $\mathcal{X}_{0}=\{\bm{0}\}$.}}}
\vspace*{-0.2in}
\label{EllipsoidalOuterApprox}
\end{figure*}

Fig. \ref{EllipsoidalOuterApprox} depicts the conservatism in ellipsoidal over-approximations $\mathcal{R}_{\text{approx}}(t)$ of the single input integrator reach sets $\mathcal{R}(\{\bm{0}\},t)\subseteq\mathcal{R}_{\text{approx}}(\{\bm{0}\},t)$ for $d=2,3,4$ with $0\leq t \leq 1$ and $\mu=1$, following the algorithms in ellipsoidal toolbox \cite{kurzhanskiy2006ellipsoidal}. Specifically, the reach set at time $t$, is over-approximated by the \emph{intersection} of a carefully constructed parameterized family of ellipsoids $\mathcal{E}\left(\bm{q}(t),\bm{Q}_{\bm{\ell}_{i}(t)}(t)\right)$ {{defined as}}
\begin{align*}
\{\bm{x}\in\mathbb{R}^{d}\mid \left(\bm{x}-\bm{q}(t)\right)\left(\bm{Q}_{\bm{\ell}_{i}(t)}(t)\right)^{-1}\left(\bm{x}-\bm{q}(t)\right)^{\top}\leq 1\},	
\end{align*}
for unit vectors $\bm{\ell}_{i}(0)\in\mathbb{R}^{d}$, $i=1,\hdots,N$. The choice of $\bm{\ell}_{i}(0)$ determines $\bm{\ell}_{i}(t) := \exp(-\bm{A}^{\top}t)\bm{\ell}_{i}(0)$, which in turn parameterizes the $d\times d$ symmetric positive definite shape matrix $\bm{Q}_{\bm{\ell}_{i}(t)}(t)$; we refer the readers to \cite[Ch. 3.2]{kurzhanskiui1997ellipsoidal}, \cite[Ch. 3]{kurzhanski2014dynamics} where the corresponding evolution equations were derived using optimal control. The center vectors $\bm{q}(t)\in\mathbb{R}^{d}$, and the shape matrices $\bm{Q}_{\bm{\ell}_{i}(t)}(t)$ for this parameterized family of ellipsoids are constructed such that $\cap_{i=1}^{N}\mathcal{E}\left(\bm{q}(t),\bm{Q}_{\bm{\ell}_{i}(t)}(t)\right)$ is guaranteed to be a superset of the reach set at time $t$ for any finite $N$, and for $N\rightarrow\infty$, recovers the reach set at that time.

For the results shown in Fig. \ref{EllipsoidalOuterApprox}, we used $N=20$ randomly chosen unit vectors $\bm{\ell}_{i}(0)\in\mathbb{R}^{d}$. Ideally, one would like to compute the (unique) minimum volume outer ellipsoid (MVOE), a.k.a. the L\"{o}wner-John ellipsoid \cite{john1948extremum,henk2012lowner} of the convex set $\cap_{i=1}^{20}\mathcal{E}\left(\bm{q}(t),\bm{Q}_{\bm{\ell}_{i}(t)}(t)\right)$, which is a semi-infinite programming problem \cite[Ch. 8.4.1]{boyd2004convex}, and has no known \emph{exact} semidefinite programming (SDP) reformulation. We computed two different relaxations of this problem: one based on the S procedure \cite[Ch. 3.7.2]{boyd1994linear}, and the other by homothetic scaling of the maximum volume inner ellipsoid (MVIE) \cite[Thm. III]{john1948extremum} of the set $\cap_{i=1}^{20}\mathcal{E}\left(\bm{q}(t),\bm{Q}_{\bm{\ell}_{i}(t)}(t)\right)$. Both of these lead to solving SDP problems, and both are guaranteed to contain the L\"{o}wner-John ellipsoid of the intersection of the parameterized family of ellipsoids. These suboptimal (w.r.t. the MVOE criterion) solutions, computed using \texttt{cvx} \cite{cvx}, are shown in Fig. \ref{EllipsoidalOuterApprox}. 

Fig. \ref{EllipsoidalOuterApprox} shows that the S procedure entail less conservatism compared to the MVIE scaling, in terms of volume. While the volume ratio trends in Fig. \ref{EllipsoidalOuterApprox} are similar to that observed in Fig. \ref{ZonotopeOuterApprox}, the approximation quality is lower. In light of the results in Sec. \ref{subsec:zonoid}, this is not surprising: the integrator reach sets being zonoids (i.e., Hausdorff limit of zonotopes), the zonotopic outer-approximations are expected to perform better than other over-approximating shape primitives. 

The main point here is that our results in Sec. \ref{sec:Size} provide the ground truth for the size of the integrator reach set, thereby help benchmarking the performance of reach set approximation algorithms.

%%%%%%%%%%%%%%%%%%%%%%%%%%%%%%%%%%%%%%%%%%%%%%%%%%%%%%%%%%%%%%%%%%%%%%%%%%%%%%%

%%%%%%%%%%%%%%%%%%%%%%%%%%%%%%%%%%%%%%%%%%%%%%%%%%%%%%%%%%%%%%%%%%%%%%%%%%%%%%%%

\section{Epilogue}\label{sec:Epilogue}
\subsubsection*{Recap}
The present paper initiates a systematic study of integrator reach set. When the input uncertainty set is hyperrectangle, we showed that the corresponding compact convex reach set $\mathcal{R}^{\Box}$ is in fact semialgebraic (Sec. \ref{subsec:semialgebraic}) as well as a zonoid (range of an atom free vector measure) up to translation (Sec. \ref{subsec:zonoid}). We derived the equation of its boundary in both parametric (Proposition \ref{Prop:ParametricRepresentationOfBoundaryPoint}) and implicit form (Sec. \ref{subsec:Implicitization}). We obtained the closed form formula for the volume (Sec. \ref{subsec:Volume}) and diameter (Sec. \ref{subsec:Diameter}) of these reach sets. We also derived the scaling laws (Sec. \ref{subsec:Scaling}) for these quantities. We pointed out that these results may be used to benchmark the performance of set over-approximation algorithms (Sec. \ref{sec:Applications}). 

\subsubsection*{What Next}
In the sequel Part II, we will show how the ideas presented herein enable computing the reach sets for feedback linearizable systems. The focus will be in computing the reach set in transformed state coordinates associated with the normal form, and to map that set back to original state coordinates under diffeomorphism. This, however, requires nontrivial extension of the basic theory presented here (especially those in Proposition \ref{Prop:ParametricRepresentationOfBoundaryPoint} and Sec \ref{subsec:Implicitization}) since we will need to handle time-dependent set-valued uncertainty in transformed control input even when the original control takes values from a set that is \emph{not} time-varying. %We will also address how to handle the state constraints. Several numerical examples will be given to illustrate the results.

%%%%%%%%%%%%%%%%%%%%%%%%%%%%%%%%%%%%%%%%%%%%%%%%%%%%%%%%%%%%%%%%%%%%%%%%%%%%%%%%

%\section*{Acknowledgement}

%%%%%%%%%%%%%%%%%%%%%%%%%%%%%%%%%%%%%%%%%%%%%%%%%%%%%%%%%%%%%%%%%%%%%%%%%%%%%%%%

\appendix

\subsection{Proof of Proposition \ref{proSptFn}}
\label{AppendixProofPro:proSptFn}
Since support function is distributive over sum, we have
\begin{align}
h_{\mathcal{R}(\mathcal{X}_{0},t)}\left(\bm{y}\right) &=  \underset{\bm{x}_{0}\in\mathcal{X}_{0}}{\sup}\langle\bm{y}_{},\exp\left(t\bm{A}_{}\right)\bm{x}_{0}\rangle \nonumber\\ 
&\quad \qquad+ h_{\int_{0}^{t}\exp(s\bm{A})\bm{B}\mathcal{U}\differential s}(\bm{y}).
\label{hRjGen}
\end{align}
The block diagonal structure of the matrix $\bm{A}$ in \eqref{defAB} implies 
\begin{align}
 \!\underset{\bm{x}_{0}\in\mathcal{X}_{0}}{\sup}\langle\bm{y}_{},\exp\left(t\bm{A}\right)\bm{x}_{0}\rangle \!=\! \underset{\bm{x}_{0}\in\mathcal{X}_{0}}{\sup} 
\sum_{j=1}^{m}\langle\bm{y}_{j},\exp\left(t\bm{A}_j\right)\bm{x}_{j0}\rangle.
\label{hRjA}
\end{align}
Following the definition of support function and \cite[Proposition 1]{haddad2020convex}, we then have
\begin{align}
\label{hrjprop}
&h_{\int_{0}^{t}\exp(s\bm{A})\bm{B}\:\mathcal{U}\:\differential s}(\bm{y}) = \int_{0}^{t} \!\!h_{\exp(s\bm{A})\bm{B}\: \mathcal{U}}\:(\bm{y})\;\differential{s}\nonumber\\ 
&\quad=\int_{0}^{t}\underset{\bm{u} \in \mathcal{U} }{\sup}~\langle \bm{y}_{},\exp(s\bm{A})\bm{B}\bm{u}\rangle~ \differential s \nonumber\\
&\quad= \int^t_0 \underset{\bm{u}\in{\rm{closure}}\left({\rm{conv}}\left(\mathcal{U}\right)\right)}{\sup}~ \sum_{j=1}^{m}~\{\langle\bm{y}_{j},\bm{\xi}_{j}(s)\rangle u_{j}\} \:\differential{s}.
\end{align}
The last equality in \eqref{hrjprop} follows from %the structures of the state and input matrices in \eqref{defAB}, wherein $\bm{\xi}$ is given in 
\eqref{xiVector}, and from the fact \cite[Prop. 6.1]{yong1999stochastic} that the reach set remains invariant under the closure of convexification of the input set $\mathcal{U}$. Substituting \eqref{hRjA} and \eqref{hrjprop} in \eqref{hRjGen} yields \eqref{SptFnIntegratorpGen}. 
\hfill\qed

\subsection{Proof of Theorem \ref{ThmSptFn}}
\label{AppendixProofThm:SptFn}
%Writing \eqref{SptFnIntegratorpGen} for the box-valued input uncertainty \eqref{BoxInputSet}, we get    
%\begin{align}
%h_{\mathcal{R}_{j}(\mathcal{X}_{j0},t)}\left(\bm{y}_{j}\right) &=\underset{\bm{x}_{0}\in\mathcal{X}_{0}}{\sup} 
%\sum_{j=1}^{m}~ \langle\bm{y}_{j},\exp\left(t\bm{A}_j\right)\bm{x}_{j0}\rangle\nonumber\\
%&+ \int^t_0 \underset{u_j\in\mathcal[\alpha_j,\beta_j] }{\sup} \langle\bm{y}_{j},\bm{\xi}_{j}(s)\rangle\: u_{j}\: \differential{s}.
%\label{hRjf}
%\end{align}
Since the uncertainties in \eqref{BoxInputSet} along different input co-ordinate axes are mutually independent, the support function of the reach set is of the form \eqref{SptFn}. Therefore, in this case, \eqref{SptFnIntegratorpGen} takes the form
\begin{align}
h_{\mathcal{R}(\mathcal{X}_{0},t)}\left(\bm{y}\right)= &\sum_{j=1}^{m}\bigg\{\underset{\bm{x}_{j0}\in\mathcal{X}_{j0}}{\sup}\langle\bm{y}_{j},\exp\left(t\bm{A}_{j}\right)\bm{x}_{j0}\rangle\nonumber\\ 
&+ \int^t_0 \underset{u_j\in\mathcal[\alpha_j,\beta_j] }{\sup} \langle\bm{y}_{j},\bm{\xi}_{j}(s)\rangle\: u_{j}\: \differential{s}\bigg\}.
\label{hRj}
\end{align}
The optimizer $u_{j}^{\text{opt}}$ of the integrand in the RHS of \eqref{hRj}, for $j\in [m]$, can be written in terms of the Heaviside unit step function $H(\cdot)$ as 
\begin{align*}
u_{j}^{\text{opt}} &= \alpha_j + (\beta_j - \alpha_j) H(\langle\bm{y}_{j},\bm{\xi}_{j}\rangle)\\
&= \alpha_j + (\beta_j - \alpha_j) \times \frac{1}{2}\left(1 + {\rm{sgn}}\left(\langle\bm{y}_{j},\bm{\xi}_{j}\rangle\right)\right),
\end{align*}
where ${\rm{sgn}}(\cdot)$ denotes the signum function. Therefore,
\begin{align}
\underset{u_j\in\mathcal[\alpha_j,\beta_j] }{\sup} \!\!\!\!\langle\bm{y}_{j},\bm{\xi}_{j}(s)\rangle u_{j} 
= \nu_{j}\langle\bm{y}_{j},\bm{\xi}_{j}(s)\rangle + \mu_{j} |\langle\bm{y}_{j},\bm{\xi}_{j}(s)\rangle|
\label{sptfnjbeforelineartransform}
\end{align}
for $0\leq s\leq t$. Substituting \eqref{sptfnjbeforelineartransform} back in (\ref{hRj}) and integrating over $s$ completes the proof.
\hfill\qed

\subsection{Proof of Theorem \ref{Thm:zonoid}}
\label{AppendixProofThm:Zonoid}
For $s\in[0,t]$, let the vector measure $\widetilde{\bm{\mu}}$ be defined as $\differential\widetilde{\bm{\mu}}(s):=\bm{\xi}(s)\differential s$ where $\bm{\xi}(s)$ is given by (\ref{xiVector}). Then $\int_{0}^{t}|\langle\bm{y},\bm{\xi}(s)\rangle|\differential s$ is exactly in the form of a support function of a zonoid (see e.g., \cite[Sec. 2]{bolker1969class}). Using the one-to-one correspondence between a compact convex set and its support function, the corresponding set is a zonoid. 

From (\ref{SptFnAffineTransform}) and (\ref{SptFnIntegratorFinal}), $\mathcal{R}^{\Box}\left(\{\bm{x}_{0}\},t\right)$ is the translation of a set with support function $\int_{0}^{t}|\langle\bm{y},\bm{\xi}(s)\rangle|\differential s$, i.e., the translation of a zonoid. Thus, $\mathcal{R}^{\Box}\left(\{\bm{x}_{0}\},t\right)$ is a zonoid.\hfill\qed

%%%%%%%%%%%%%%%%%%%%%%%%%%%%%%%%%%%%%%%%%%%%%%%%%%%%%%%%%%%%%%%%%%%%%%%%%%%%%%%%

\subsection{Proof of Proposition \ref{Prop:ParametricRepresentationOfBoundaryPoint}}\label{AppendixProofProp:ParametricRepresentationOfBoundaryPoint}
From Sec. \ref{subsubsec:sptfn}, the supporting hyperplane at any $\bm{x}^{\textup{bdy}}\in\partial\mathcal{R}^{\Box}\left(\{\bm{x}_{0}\},t\right)$ with outward normal $\bm{y}\in\mathbb{R}^{d}$ is $\langle\bm{y},\bm{x}^{\textup{bdy}}\rangle = h_{\mathcal{R}^{\Box}\left(\{\bm{x}_{0}\},t\right)}(\bm{y})$, and the Legendre-Fenchel conjugate 
\begin{align}
h_{\mathcal{R}^{\Box}\left(\{\bm{x}_{0}\},t\right)}^{*}\left(\bm{x}^{\textup{bdy}}\right) = 0.
\label{BiconjugateEqualsToZero}	
\end{align} 

For $j\in[m]$, let $\bm{y}$ comprise of subvectors $\bm{y}_{j}\in\mathbb{R}^{r_{j}}$. Since the Cartesian product (\ref{CartesianProduct}) is equivalent to the Minkowski sum $\mathcal{R}_{1}\dotplus \hdots \dotplus\mathcal{R}_{m}$, and the support function of Minkowski sum is the sum of support functions of the summand sets \cite[p. 48]{schneider2014convex}, we have
\begin{align}
h_{\mathcal{R}^{\Box}\left(\{\bm{x}_{0}\},t\right)}(\bm{y}) &= \sum_{j=1}^{m} h_{\mathcal{R}_{j}\left(\{\bm{x}_{0}\},t\right)}(\bm{y}_{j})\nonumber\\
\Rightarrow h_{\mathcal{R}^{\Box}\left(\{\bm{x}_{0}\},t\right)}^{*}\left(\bm{x}^{\textup{bdy}}\right) &= \sum_{j=1}^{m} h_{\mathcal{R}_{j}\left(\{\bm{x}_{0}\},t\right)}^{*}\left(\bm{x}^{\textup{bdy}}_{j}\right),
\label{ConjugateOfSeparableSum}	
\end{align}
wherein the last line follows from the property that the Legendre-Fenchel conjugate of a \emph{separable} sum equals to the sum of the Legendre-Fenchel conjugates \cite[p. 95]{boyd2004convex}.

Therefore, combining (\ref{BiconjugateEqualsToZero}) and (\ref{ConjugateOfSeparableSum}), we obtain
%\begin{align}
%\underset{\bm{y}\in\mathbb{R}^{d}}{\inf}\bigg\{\langle-\bm{x}^{\textup{bdy}}+\exp\left(t\bm{A}\right)\bm{x}_{0},\bm{y}\rangle \!+ \!\!\int_{0}^{t}\!\!|\langle\bm{y},\bm{\xi}(s)\rangle| \:\differential s\bigg
%\} = 0.
%\label{convertingsuptoinf}	
%\end{align}
%For $j=1,\hdots,m$, let $\bm{y}$ comprise of subvectors $\bm{y}_{j}\in\mathbb{R}^{r_{j}}$, and rewrite (\ref{convertingsuptoinf}) as the unconstrained minimization of a separable sum
%\[\underset{\bm{y}\in\mathbb{R}^{d}}{\inf}\sum_{j=1}^{m}\langle-\bm{x}_{j}^{\textup{bdy}}+\exp\left(t\bm{A}_{j}\right)\bm{x}_{j0},\bm{y}_{j}\rangle + \mu_{j}\!\!\int_{0}^{t}\!\!|\langle\bm{y}_{j},\bm{\xi}_{j}(s)\rangle| \:\differential s = 0.\]
%Hnece the minimizer $\bm{y}^{\text{opt}}$ comprises of subvectors $\bm{y}_{j}^{\text{opt}}\in\mathbb{R}^{r_{j}}$ solving
\begin{align}
&\sum_{j=1}^{m}\:\underset{\bm{y}_{j}\in\mathbb{R}^{r_{j}}}{\inf}\!\bigg\{\!\langle-\bm{x}^{\textup{bdy}}_{j}+\exp\left(t\bm{A}_{j}\right)\bm{x}_{j0}+\nu_{j}\bm{\zeta}_j(t),\bm{y}_{j}\rangle \nonumber\\
&\qquad\qquad\qquad\qquad\qquad + \mu_{j}\!\!\int_{0}^{t}\!\!|\langle\bm{y}_{j},\bm{\xi}_{j}(s)\rangle| \:\differential s\!\bigg
\} = 0.
\label{UnconstrainedMinimizationSeparableSum}	
\end{align}
For $j\in[m]$, since each objective in (\ref{UnconstrainedMinimizationSeparableSum}) involves an integral of the absolute value of a polynomial in $s$ of degree $r_{j}-1$, that polynomial can have \emph{at most} $r_{j}-1$ roots in the interval $[0,t]$, i.e., can have at most $r_{j}-1$ sign changes in that interval. If all $r_{j}-1$ roots of the aforesaid polynomial are in $[0,t]$, we denote these roots as $s_{1} \leq s_{2} \leq \hdots \leq s_{r_{j}-1}$, and write
\begin{align}
&\int_{0}^{t}\!\!|\langle\bm{y}_{j},\bm{\xi}_{j}(s)\rangle|\:\differential s = \pm\int_{0}^{s_{1}}\!\!\langle \bm{y}_{j},\bm{\xi}_{j}(s)\rangle\:\differential s \mp \int_{s_{1}}^{s_{2}}\!\!\langle \bm{y}_{j},\bm{\xi}_{j}(s)\rangle\:\differential s\nonumber\\
&\qquad\qquad\qquad\qquad\quad\pm\hdots \pm (-1)^{r_{j}-1}\int_{s_{r_{j}-1}}^{t}\!\!\langle \bm{y}_{j},\bm{\xi}_{j}(s)\rangle\:\differential s\nonumber\\
&= \langle\bm{y}_{j}, \pm\bm{\zeta}_{j}(0,s_1)\mp\bm{\zeta}_{j}(s_1,s_2)\pm\hdots \pm (-1)^{r_{j}-1}\bm{\zeta}_{j}(s_{r_{j}-1},t)\rangle.
\label{SumOfIntegrals}
\end{align}
Notice that even if the number of roots in $[0,t]$ is strictly less than\footnote{this may happen either because there are repeated roots in $[0,t]$, or because some real roots exist outside $[0,t]$, or because some roots are complex conjugates, or a combination of the previous three.} $r_{j}-1$, the expression (\ref{SumOfIntegrals}) is generic in the sense the corresponding summand integrals become zero. Thus, combining (\ref{UnconstrainedMinimizationSeparableSum}) and (\ref{SumOfIntegrals}), we arrive at
\begin{align}
&\sum_{j=1}^{m}\:\underset{\bm{y}_{j}\in\mathbb{R}^{r_{j}}}{\inf}\langle-\bm{x}^{\textup{bdy}}_{j}+\exp\left(t\bm{A}_{j}\right)\bm{x}_{j0} +\nu_{j}\bm{\zeta}_j(t) \pm\mu_{j}\bm{\zeta}_{j}(0,s_1)\nonumber\\
& \mp\mu_{j}\bm{\zeta}_{j}(s_1,s_2)\pm\hdots\pm\mu_{j} (-1)^{r_{j}-1}\bm{\zeta}_{j}(s_{r_{j}-1},t),\bm{y}_{j}\rangle = 0.
\label{LinearMinimizationyj}	
\end{align}

The left hand side of (\ref{LinearMinimizationyj}) being the sum of the infimum values of linear functions, can achieve zero if and only if each of those infimum equals to zero, i.e., if and only if
\begin{align}
&\bm{x}^{\textup{bdy}}_{j}=\exp\left(t\bm{A}_{j}\right)\bm{x}_{j0} +\nu_{j}\bm{\zeta}_j(t) \pm\mu_{j}\bm{\zeta}_{j}(0,s_1)\mp\mu_{j}\bm{\zeta}_{j}(s_1,s_2)\nonumber\\
&\qquad\quad\pm\hdots\pm (-1)^{r_{j}-1}\mu_{j}\bm{\zeta}_{j}(s_{r_{j}-1},t).
\label{CoeffEqualToZero}	
\end{align}
Using (\ref{diagBlocksOfSTM}), (\ref{xiVector}) and (\ref{DefzetaDoubleArgument}), we simplify (\ref{CoeffEqualToZero}) to (\ref{ParamRepresentationBoundary}), thereby completing the proof. \hfill\qed

%%%%%%%%%%%%%%%%%%%%%%%%%%%%%%%%%%%%%%%%%%%%%%%%%%%%%%%%%%%%%%%%%%%%%%%%%%%%%%%%

\subsection{Proof of Corollary \ref{Corollary:TwoBoundingSurfaces}}\label{AppendixProofCorr:TwoBoundingSurfaces}
From (\ref{ParamRepresentationBoundary}), we get two different parametric representations of $\bm{x}_{j}^{\text{bdy}}$ in terms of $(s_{1}, s_{2}, \hdots, s_{r_{j}-1})$. One parametric representation results from the choice of positive sign for the $\pm$ appearing in (\ref{ParamRepresentationBoundary}), and another for the choice of negative sign for the same. Denoting the implicit representation corresponding to the parametric representation (\ref{ParamRepresentationBoundary}) with $+$ (resp. $-$) sign as $p_{j}^{\textup{upper}}(\bm{x})=0$ (resp. $p_{j}^{\textup{lower}}(\bm{x})=0$), the result follows. \hfill\qed

%%%%%%%%%%%%%%%%%%%%%%%%%%%%%%%%%%%%%%%%%%%%%%%%%%%%%%%%%%%%%%%%%%%%%%%%%%%%%%%%

\subsection{Proof of Theorem \ref{Thm:semialgebraic}}\label{AppendixProofThm:semialgebraic}
We notice that (\ref{ParamRepresentationBoundary}) gives polynomial parameterizations of the components of $\bm{x}_{j}^{\text{bdy}}$ for all $j\in[m]$. In particular, for each $k\in[r_{j}]$, the right hand side of (\ref{ParamRepresentationBoundary}) is a homogeneous polynomial in $r_{j}-1$ parameters $(s_{1},s_{2},\hdots,s_{r_{j}-1})$ of degree $r_{j}-k+1$. By polynomial implicitization \cite[p. 134]{cox2013ideals}, the corresponding implicit equations $p_{j}^{\text{upper}}\left(\bm{x}_{j}^{\text{bdy}}\right)=0$ (when fixing plus sign for $\pm$ in (\ref{ParamRepresentationBoundary})) and $p_{j}^{\text{lower}}\left(\bm{x}_{j}^{\text{bdy}}\right)=0$ (when fixing minus sign for $\pm$ in (\ref{ParamRepresentationBoundary})), must define affine varieties $V_{\mathbb{R}[x_1,...,x_{r_{j}}]}(p_{j}^{\text{upper}}), V_{\mathbb{R}[x_1,...,x_{r_{j}}]}(p_{j}^{\text{lower}})$ in $\mathbb{R}\left[x_1,\hdots,x_d\right]$.

Specifically, denote the right hand sides of (\ref{ParamRepresentationBoundary}) as $g_{1}^{\pm},\hdots,g_{r_{j}}^{\pm}$ for all $j\in[m]$, where the superscripts indicate that either all $g$'s are chosen with plus signs, or all with minus signs. Then write (\ref{ParamRepresentationBoundary}) as
\begin{align*}
\bm{x}^{\text{bdy}}_{j}(1) &= g_{1}^{\pm}\left(s_{1},s_{2},\hdots,s_{r_{j}-1}\right),\\
%\bm{x}^{\text{bdy}}_{j}(2) &= g_{2}^{\pm}\left(s_{1},s_{2},\hdots,s_{r_{j}-1}\right),\\
 & \;\;\vdots\\
\bm{x}^{\text{bdy}}_{j}(r_{j}) &= g_{r_{j}}^{\pm}\left(s_{1},s_{2},\hdots,s_{r_{j}-1}\right).	
\end{align*}
Now for each $j\in[m]$, consider the ideal 
\begin{align*}
I_{j}^{\pm} := &\langle\langle\bm{x}^{\text{bdy}}_{j}(1) - g_{1}^{\pm}, \bm{x}^{\text{bdy}}_{j}(2) - g_{2}^{\pm},\hdots,\bm{x}^{\text{bdy}}_{j}(r_{j}) - g_{r_{j}}^{\pm}\rangle\rangle\\
&\subseteq \mathbb{R}[s_{1},s_{2},\hdots,s_{r_{j}-1},x_1,x_{2},\hdots,x_{r_{j}}],	
\end{align*}
and let $I_{j,r_{j}-1}^{\pm}:=I_{j}^{\pm}\cap\mathbb{R}[x_1,...,x_{r_{j}}]$ be the $(r_{j}-1)$th elimination ideal of $I_{j}^{\pm}$. Then for each $j\in[m]$, the variety \[V\left(I_{j,r_{j}-1}^{+}\right) = V_{\mathbb{R}[x_1,...,x_{r_{j}}]}(p_{j}^{\text{upper}}).\] Likewise, the variety $V\left(I_{j,r_{j}-1}^{-}\right) = V_{\mathbb{R}[x_1,...,x_{r_{j}}]}(p_{j}^{\text{lower}})$. 

Thus, the algebraic boundary (i.e., the Zariski closure of the Euclidean boundary) of $\mathcal{R}_{j}$ is
\[\partial\mathcal{R}_{j} = V_{\mathbb{R}[x_1,...,x_{r_{j}}]}\left(p_{j}^{\text{upper}}\right) \cup V_{\mathbb{R}[x_1,...,x_{r_{j}}]}\left(p_{j}^{\text{lower}}\right).\]
Therefore, $\mathcal{R}_{j} := \{\bm{x}\in\mathbb{R}^{r_{j}}\mid p_{j}^{\textup{upper}}(\bm{x})\leq 0, \; p_{j}^{\textup{lower}}(\bm{x})\leq 0\}$ is semialgebraic for all $j\in[m]$.

Since the Cartesian product of semialgebraic sets is semialgebraic, the statement follows from (\ref{CartesianProduct}).\hfill\qed

%%%%%%%%%%%%%%%%%%%%%%%%%%%%%%%%%%%%%%%%%%%%%%%%%%%%%%%%%%%%%%%%%%%%%%%%%%%%%

\subsection{Proof of Theorem \ref{ThmVolIntegratorReachSet}}\label{AppendixProofThm:volume}
We organize the proof in three steps.\\
\noindent \ul{Step 1:} From (\ref{CartesianProduct}), we have
\begin{align}
\vol \left(\mathcal{R}^{\Box}\left(\{\bm{x}_{0}\},t\right)\right) &= \vol \left(\mathcal{R}_{1}\times\mathcal{R}_{2}\times\hdots\times\mathcal{R}_{m}\right)\nonumber\\
&=\displaystyle\prod_{j=1}^{m} \vol\left(\mathcal{R}_{j}\left(\{\bm{x}_{0}\},t\right)\right).
\label{VolCartesianProduct}	
\end{align}
\noindent \ul{Step 2:} Motivated by (\ref{VolCartesianProduct}), we focus on deriving the $r_{j}$-dimensional volume of $\mathcal{R}_{j}\left(\{\bm{x}_{0}\},t\right)$. For this purpose, we proceed as in \cite{haddad2020convex} by uniformly discretizing the interval $[0,t]$ into $n$ subintervals $[(i-1)t/n,it/n)$, $i=1,\hdots,n$,
%\[\bigg[\frac{(i-1)t}{n}, \frac{it}{n}\bigg), \quad i=1, 2, \hdots, n,\]
with $(n+1)$ breakpoints $\{t_{i}\}_{i=0}^{n}$, where $t_{i}:=it/n$ for $i=0,1,\hdots,n$.

%From (\ref{blkdiagAB}) and (\ref{SetValuedIntegral}), we then have
From (\ref{LimitInterpretationZonoid}), we then have   
\begin{align}
%&\vol \left(\mathcal{R}_{j}\left(\{\bm{x}_{0}\},t\right)\right) =\mathrm{vol}\left(\int_{0}^{t}\!\!\!\exp\left(s\bm{A}_{j}\right)\bm{b}_{j} [-\mu_{j}, \mu_{j}] \differential s\right)\nonumber \\
%\newline \nonumber \\
&\vol \left(\mathcal{R}_{j}\left(\{\bm{x}_{0}\},t\right)\right) = \vol\left(\displaystyle\lim_{n\rightarrow\infty} \displaystyle\sum_{i=0}^{n}\frac{t}{n}\exp\left(t_{i}\bm{A}_{j}\right)\bm{b}_{j} [-\mu_{j}, \mu_{j}] \right)\nonumber \\
\newline \nonumber\\
&= \!\displaystyle\lim_{n\rightarrow\infty}\!\left(\dfrac{t}{n}\right)^{\!\!r_{j}}\!\!\vol\left(\displaystyle\sum_{i=0}^{n}\exp\left(t_{i}\bm{A}_{j}\right)\bm{b}_{j} [-\mu_{j}, \mu_{j}]\right)\nonumber \\
\newline \nonumber\\
&=t^{r_{j}}\displaystyle\lim_{n\rightarrow\infty}\dfrac{1}{n^{r_{j}}}\vol \left(\displaystyle\sum_{i=0}^{n}\mu_j\bm{\xi}_j(t_i)[-1,1]\right),
\label{Volintermed}
\end{align}
where $\bm{\xi}_j$ was defined in (\ref{xiVector}).	We recognize that the set $\sum_{i=0}^{n}\mu_j \bm{\xi}_j(t_i)[-1,1]$
in (\ref{Volintermed}) is a Minkowski sum of $n+1$ intervals, i.e., is a zonotope imbedded in $\mathbb{R}^{r_{j}}$, wherein each interval is rotated and scaled in $\mathbb{R}^{r_{j}}$ via different linear transformations $\exp(t_i\bm{A}_{j})$, $i=0,1,\hdots,n$. %In other words, the set (\ref{DefSet}) is a zonotope imbedded in $\mathbb{R}^{r_{j}}$.

Using the formula for the volume of zonotopes \cite[eqn. (57)]{shephard1974combinatorial}, \cite[exercise 7.19]{ziegler2012lectures}, we can write (\ref{Volintermed}) as
\begin{align}
\mathrm{vol}&\left(\mathcal{R}_{j}\left(\{\bm{x}_{0}\},t\right)\right) = (2\mu_{j}t)^{r_{j}}\displaystyle\lim_{n\rightarrow\infty}\dfrac{1}{n^{r_{j}}}\:\times\nonumber\\
&\displaystyle\sum_{0\leq i_{1}< i_{2} < \hdots < i_{r_{j}} \leq n}\!\!\!\!\det\left(\bm{\xi}_{j}(t_{i_{1}}) | \bm{\xi}_{j}(t_{i_{2}}) | \hdots | \bm{\xi}_{j}(t_{i_{r_{j}}})\right).
\label{VolRj}
\end{align}
To compute the summand determinants in (\ref{VolRj}), let
\[\Delta_{j}\left(i_{1}, i_{2}, \hdots, i_{r_{j}}\right) := \det\left(\bm{\xi}_{j}(t_{i_{1}}) | \bm{\xi}_{j}(t_{i_{2}}) | \hdots | \bm{\xi}_{j}(t_{i_{r_{j}}})\right),\]
where $0\leq i_{1}< i_{2} < \hdots < i_{r_{j}} \leq n$. In the matrix list notation, let us use the vertical bars $|\cdot|$ to denote the absolute value of determinant. From (\ref{xiVector}), $\Delta_{j}\left(i_{1}, i_{2}, \hdots, i_{r_{j}}\right)$ equals
{\small{\begin{align}\allowbreak
&\begin{vmatrix}
 \dfrac{\left(i_{1}t/n\right)^{r_{j}-1}}{(r_{j}-1)!} & \dfrac{\left(i_{2}t/n\right)^{r_{j}-1}}{(r_{j}-1)!} & \hdots & \dfrac{\left(i_{r_{j}}t/n\right)^{r_{j}-1}}{(r_{j}-1)!}\\
  & & &\\
  \dfrac{\left(i_{1}t/n\right)^{r_{j}-2}}{(r_{j}-2)!} & \dfrac{\left(i_{2}t/n\right)^{r_{j}-2}}{(r_{j}-2)!} & \hdots & \dfrac{\left(i_{r_{j}}t/n\right)^{r_{j}-2}}{(r_{j}-2)!}\\ 
   &&&\\ 
 \vdots & \vdots & \vdots & \vdots\\
  &&&\\
 i_{1}t/n & i_{2}t/n & \hdots & i_{r_{j}}t/n\\
 &&&\\
 1 & 1 & \hdots & 1
 \end{vmatrix}\nonumber\\
\allowbreak &= \dfrac{(t/n)^{1 + 2 + \hdots + (r_{j}-1)}}{1!\times2!\times\hdots\times(r_{j}-1)! }\begin{vmatrix}
i_{1}^{r_{j}-1} & i_{2}^{r_{j}-1} & \hdots & i_{r_{j}}^{r_{j}-1}\\
  & & &\\
 i_{1}^{r_{j}-2} & i_{2}^{r_{j}-2} & \hdots & i_{r_{j}}^{r_{j}-2}\\ 
   &&&\\ 
 \vdots & \vdots & \vdots & \vdots\\
  &&&\\
 i_{1} & i_{2} & \hdots & i_{r_{j}}\\
 &&&\\
 1 & 1 & \hdots & 1
 \end{vmatrix}, \nonumber\\
&=  \dfrac{(t/n)^{r_{j}(r_{j}-1)/2}}{\displaystyle\prod_{k=1}^{r_{j}-1}k!}\begin{vmatrix}
 1 & 1 & \hdots & 1\\
  & & &\\
   i_{1} & i_{2} & \hdots & i_{r_{j}}\\ 
   &&&\\ 
 \vdots & \vdots & \vdots & \vdots\\
  &&&\\  
 i_{1}^{r_{j}-2} & i_{2}^{r_{j}-2} & \hdots & i_{r_{j}}^{r_{j}-2}\\
 &&&\\
i_{1}^{r_{j}-1} & i_{2}^{r_{j}-1} & \hdots & i_{r_{j}}^{r_{j}-1}\\
 \end{vmatrix},
 \label{DetStructure}	
\end{align}}}
where we used the properties of elementary row operations. 

Notice that the determinant appearing in the last step of (\ref{DetStructure}) is the \emph{Vandermonde determinant} (see e.g., \cite[p. 37]{horn2012matrix})
\begin{align}
\displaystyle\prod_{1 \leq a < b \leq r_{j}} \left(i_{b} - i_{a}\right).
\label{VandermondeDet}	
\end{align}
Combining (\ref{VolRj}), (\ref{DetStructure}) and (\ref{VandermondeDet}), we obtain
\begin{align}
\mathrm{vol}&\left(\mathcal{R}_{j}\left(\{\bm{x}_{0}\},t\right)\right) = \dfrac{(2\mu_{j})^{r_{j}}t^{r_{j}(r_{j}+1)/2}}{\displaystyle\prod_{k=1}^{r_{j}-1}k!}\displaystyle\lim_{n\rightarrow\infty}\dfrac{1}{n^{r_{j}(r_{j}+1)/2}}\:\times\nonumber\\
&\displaystyle\sum_{0\leq i_{1}< i_{2} < \hdots < i_{r_{j}} \leq n}\quad\displaystyle\prod_{1 \leq a < b \leq r_{j}} \left(i_{b} - i_{a}\right).
\label{VolWithLimAndNestedSum}	
\end{align}
\ul{Step 3:} Our next task is to simplify (\ref{VolWithLimAndNestedSum}). %by eliminating the limit and the nested sums. 
Observe that the sum 
\begin{align}
\displaystyle\sum_{0\leq i_{1}< i_{2} < \hdots < i_{r_{j}} \leq n}\quad\displaystyle\prod_{1 \leq a < b \leq r_{j}} \left(i_{b} - i_{a}\right),
\label{SumPoly}	
\end{align}
returns a polynomial in $n$ of degree $r_{j}(r_{j}+1)/2$, and hence the limit in (\ref{VolWithLimAndNestedSum}) is always well-defined. Specifically, the limit extracts the leading coefficient of this polynomial. 

Let us denote the leading coefficient of the sum (\ref{SumPoly}) as $c(r_{j})$. By the Euler-Maclaurin formula \cite{apostol1999elementary}, \cite[Chap. II.10]{hairer2006analysis}:
\begin{align}
\!\!\!\!c(r_{j}) = \!\!\!\displaystyle\int\limits_{0\leq y_{1} < y_{2} < \hdots < y_{r_{j}} \leq 1}\;\prod\limits_{1 \le \alpha < \beta \le r_{j}}\!\!(y_a - y_b)\cdot  \prod\limits_{a=1}^{r_{j}} \differential y_a.
\label{EulerMaclaurin}	
\end{align}
One way to unpack (\ref{EulerMaclaurin}) is to write it as a sum over the symmetric permutation group $\mathfrak{S}_{r_{j}}$ of the finite set $[r_{j}]$, i.e.,
\[\!\!c(r_{j}) =\displaystyle\sum_{\sigma\in \mathfrak{S}_{r_{j}}}\text{sgn}(\sigma)\displaystyle\frac{1}{\prod_{k=1}^{r_{j}}(\sigma_{1} + \sigma_{2} + ... + \sigma_{k})},\]
where $\text{sgn}(\sigma) := (-1)^{\nu}$, $\nu := \{\#(i,j) \mid i<j, \sigma(i)>\sigma(j)\}$, and $\#$ stands for ``the number of". We will now prove that
\begin{align}
c(r_{j}) = \displaystyle\prod_{k=1}^{r_{j}-1} \dfrac{\left(k!\right)^{2}}{(2k+1)!}.
\label{crjClosedForm}	
\end{align}
To this end, we write $r_{j}!\cdot c(r_{j})$ as an integral over $[0,1]^{r_{j}}$:
\begin{align}
r_{j}!\cdot c(r_{j}) = \int_{[0,1]^{r_{j}}}\prod_{1\leq a < b \leq r_{j}} |y_{a} - y_{b}| \: {\rm{d}}y_{1} ... {\rm{d}}y_{r_{j}}.
\label{IntegralOverUnitCube}	
\end{align}
In 1955, de Bruijn \cite[see toward the end of Sec. 9]{de1955some} used certain Pfaffians to evaluate
\begin{align*}
&\int_{[0,1]^{r_{j}}}\prod_{1\leq a < b \leq r_{j}} |y_{a} - y_{b}| \: {\rm{d}}y_{1} ... {\rm{d}}y_{r_{j}}\\
=& \frac{r_{j}!\cdot\{1! \times 2! \times ... \times (r_{j}-1)!\}^{2}}{1!\times 3!\times ... \times (2r_{j}-1)!}, \quad r_{j}=2,3,\hdots,
\end{align*}
which upon substitution in (\ref{IntegralOverUnitCube}), indeed yields (\ref{crjClosedForm}).

Combining (\ref{VolWithLimAndNestedSum}) and (\ref{crjClosedForm}), we arrive at
\begin{align}
\!\!\!\mathrm{vol}\left(\mathcal{R}_{j}\left(\{\bm{x}_{0}\},t\right)\right) &= \dfrac{(2\mu_{j})^{r_{j}}t^{r_{j}(r_{j}+1)/2}}{\displaystyle\prod_{k=1}^{r_{j}-1}k!} c(r_{j})\nonumber\\
&= (2\mu_{j})^{r_{j}}t^{r_{j}(r_{j}+1)/2} \displaystyle\prod_{k=1}^{r_{j}-1} \dfrac{k!}{(2k+1)!}.
\label{volRjclosedform}	
\end{align}
Finally, substituting (\ref{volRjclosedform}) in (\ref{VolCartesianProduct}), and recalling that $r_{1}+r_{2}+\hdots + r_{m} = d$, the expression (\ref{VolumeFormula}) follows.\hfill\qed

%%%%%%%%%%%%%%%%%%%%%%%%%%%%%%%%%%%%%%%%%%%%%%%%%%%%%%%%%%%%%%%%%%%%%%%%%%%%%

\subsection{Proof of Theorem \ref{ThmDiamIntegratorReachSet}}\label{AppendixProofThm:diam}
From (\ref{xiVector}), the subvector $\bm{\xi}_{j}(s)$, where $j\in[m]$, is component-wise nonnegative for all $s\in[0,t]$. 

Therefore, by triangle inequality, we have
\begin{align}
\!\int_{0}^{t}\!\lvert\langle \bm{\eta},\bm{\xi}(s)\rangle\rvert\:\differential s \leq \!\int_{0}^{t}\!\sum_{j=1}^{m} \langle\lvert\bm{\eta}_{j}\rvert,\mu_{j}\bm{\xi}_{j}(s)\rangle = \langle|\bm{\eta}|,\bm{\zeta}(t)\rangle,
\label{TrinagleInequality}	
\end{align}
where $|\bm{\eta}_{j}|$ denotes the $j$th subvector with component-wise absolute values. Let us call $|\bm{\eta}|$ as the ``absolute unit vector".

The upper bound in (\ref{TrinagleInequality}) is convex in $\bm{\eta}$, and is maximized by an absolute unit vector collinear with $\bm{\zeta}(t)$ given by%i.e., or equivalently by $\bm{\eta}\in\mathbb{S}^{d-1}$ such that 
 \begin{align}
\bm{\eta} = \pm \dfrac{\bm{\zeta}(t)}{\parallel\bm{\zeta}(t)\parallel_{2}},
\label{MaximizerOfUpperBound}	
\end{align}
i.e., the unit vectors associated with $\bm{\zeta}(t)$ up to plus-minus sign permutations among its components. 

Out of the $2^{d}$ unit vectors given by (\ref{MaximizerOfUpperBound}), the ``all plus" and ``all minus" unit vectors achieve equality in (\ref{TrinagleInequality}), and hence must be the maximizers of (\ref{widthformula}). The inequality (\ref{TrinagleInequality}) remains strict for the remaining $2^{d}-2$ unit vectors in (\ref{MaximizerOfUpperBound}), thus are suboptimal for (\ref{widthformula}). Therefore, the maximizers in (\ref{DefDiam}) are
\[\bm{\eta}^{\max} = \bm{\zeta}(t)/\parallel\bm{\zeta}(t)\parallel_{2}, \quad -\bm{\zeta}(t)/\parallel\bm{\zeta}(t)\parallel_{2},\]
which upon substitution in (\ref{widthformula}), results in (\ref{expandedformDiamFormula}).\hfill\qed

%%%%%%%%%%%%%%%%%%%%%%%%%%%%%%%%%%%%%%%%%%%%%%%%%%%%%%%%%%%%%%%%%%%%%%%%%%%%%%%%
\bibliographystyle{IEEEtran}
\bibliography{references.bib}

\balance

%\begin{IEEEbiography}[{\includegraphics[height=1.18in]{Shadi.jpg}}]{Shadi
%Haddad}
%is a Ph.D. student in the Department of Applied Mathematics at
%University of California, Santa Cruz. She received M.S. in Mechanical Engineering from the University of Tehran in 2018. Her research
%focus is on control and optimization.
%\end{IEEEbiography}
%
%
%\begin{IEEEbiography}[{\includegraphics[height=1.25in]{Abhishek-Halder.jpg}}]{Abhishek Halder} 
%is an Assistant Professor in the Department of Applied Mathematics, and an affiliated faculty in the Department of Electrical and Computer Engineering at University of California, Santa Cruz. Before that he held postdoctoral positions in the Department of Mechanical and Aerospace Engineering at University of California, Irvine, and in the Department of Electrical and Computer Engineering at Texas A\&M University. He obtained his Bachelors and Masters from Indian Institute of Technology Kharagpur in 2008, and Ph.D. from Texas A\&M University in 2014, all in Aerospace Engineering. His research interests are in stochastic systems, control and optimization with application focus on large scale cyber-physical systems.
%\end{IEEEbiography}

\end{document}